\documentclass[10pt,aps,twocolumn,prd,superscriptaddress,showpacs,nofootinbib,noshowkeys,floatfix,showpacs,preprintnumbers]{revtex4-1}
\usepackage{graphicx}
\usepackage{amsmath, amssymb}
\usepackage{multirow}
\usepackage{longtable}
\begin{document}
\title{Probing deconfinement in a chiral effective model with Polyakov loop at 
imaginary chemical potential}
\date{\today}
\author{Kenji Morita}
\email{kmorita@yukawa.kyoto-u.ac.jp}
\affiliation{GSI, Helmholzzentrum f\"{u}r Schwerionenforschung,
Planckstr. 1, D-64291 Darmstadt, Germany}
\affiliation{Yukawa Institute for Theoretical Physics, Kyoto University, Kyoto 606-8502, Japan}
\author{Vladimir Skokov}
\author{Bengt Friman}
\affiliation{GSI, Helmholzzentrum f\"{u}r Schwerionenforschung,
Planckstr. 1, D-64291 Darmstadt, Germany}
\author{Krzysztof Redlich}
\affiliation{Institute of Theoretical Physics, University of Wroclaw,
PL-50204 Wroc\l aw, Poland}
\affiliation{Extreme Matter Institute EMMI, GSI, 
Planckstr. 1, D-64291 Darmstadt, Germany}
\preprint{YITP-11-70}
\begin{abstract}
 The phase structure of the two-flavor Polyakov-loop extended
 Nambu-Jona-Lashinio model is explored at finite temperature and
 imaginary chemical potential with a particular emphasis on the
 confinement-deconfinement transition. We point out that the confined
 phase is characterized by  a $\cos3\mu_I/T$ dependence of the chiral
 condensate on the  imaginary chemical potential 
 while in the deconfined phase this dependence is given by $\cos\mu_I/T$ and accompanied by a cusp structure induced by the $Z(3)$ transition.
 We demonstrate that the phase structure of the model strongly depends
 on the choice of the Polyakov loop potential $\mathcal{U}$. Furthermore, we find that
 by changing the four fermion coupling constant $G_s$, the
 location of the critical endpoint of the deconfinement transition can
 be moved into the real chemical potential region.
 We propose a new parameter characterizing the
 confinement-deconfinement transition.
\end{abstract}
\maketitle

\section{Introduction}

The exploration of the phase diagram of strongly interacting matter has
received a lot of attention in recent years. 
First principle calculations of the phase structure from the Lagrangian
of  Quantum Chromodynamics (QCD) is intrinsically difficult owing to the
strongly coupled nature of the theory at large distances.
Lattice Gause Theory (LGT) calculations provide a unique and powerful 
tool for studying QCD in the non-perturbative regime.  
Increasing computer power has recently made LGT simulations at almost
physical quark masses possible \cite{bazavov09:_equat_qcd,aoki09:_qcd}. 
The results indicate that the transition
from the confined, chirally broken phase to the deconfined, chirally
restored phase at $T\sim 160$
MeV~ and vanishing baryon
chemical potential, $\mu_{B}=0$, is of the crossover type \cite{fodor}.

For non-zero net baryon density, LGT calculations suffer from the
so-called ``sign problem''.  For finite quark chemical potential (and
$N_c=3$), the statistical weight of the Monte-Carlo simulation becomes
non positive definite due to the complex fermion determinant. This issue
has impeded the progress in LGT calculations at finite densities.  

There have been several attempts to bypass the sign problem
\cite{muroya03:_lattic_qcd}. One interesting approach involves using an
imaginary chemical potential, for which the fermion determinant is real
and, therefore, systematic LGT simulations are
possible~\cite{roberge86:_gauge_qcd}. 
There are two major ways for extracting information on the real phase diagram 
from calculations at imaginary chemical potential. One is to project 
the grand partition function $Z_G$ computed at imaginary chemical potential
onto the canonical partition function 
\begin{equation}
 Z_c(T,V,N_q) = \intop_{0}^{2\pi}\frac{d(\beta\mu_I)}{2\pi}e^{-i\beta\mu_I N_q}Z_G(T,V,\mu_q=i\mu_I).\label{eq:canonical}
\end{equation}
In spite of the difficulties involved in the evaluation of the oscillatory integral, there are lattice calculations aimed at
studying the phase diagram in the temperature-number density plane by means 
of this approach~\cite{ejiri08:_canon_qcd,li:_finit_qcd_n_n}. An
alternative way involves an analytic continuation from imaginary to real
values of the chemical potential. This method has proven quite powerful
for determining the critical line at  $\mu_q < \pi T/3$
\cite{forcrand02:_qcd} and this approach has been
applied in LGT calculations 
\cite{forcrand03:_qcd,d'elia03:_finit_qcd,Papa,d'elia04:_qcd_b,chen05:_phase_qcd_wilson,d'elia07:_stron_qcd,forcrand07:_n_qcd,wu07:_phase_qcd_wilson,
forcrand:_const_qcd,nagata11:_imagin_chemic_poten_approac_for}, 
in resummed perturbation theory \cite{hart01:_testing}, as well as in
quasiparticle models \cite{bluhm08:_quasiparticle}.
Of course, the analytic continuation requires
knowledge of the analytic structure of the thermodynamic functions. 
Therefore effective models that share the symmetries of QCD
are useful for testing such approaches, 
because a result obtained by analytic continuation
from imaginary chemical potential can be confronted with the known
solution at real chemical potential. 

A remarkable feature of QCD at imaginary chemical potential is the
Roberge-Weiss (RW) transition at $\mu_q/T = \pi/3+2\pi k/3$, where $k$
is an integer \cite{roberge86:_gauge_qcd}. The RW transition involves a
shift from one $Z(3)$ sector to another in the deconfined phase. 
This transition is a remnant of the $Z(3)$ symmetry
of the pure gauge theory, which is explicitly broken in
the presence of fermions of finite mass. Note that in this case the Polyakov loop is not an exact
order parameter.
Using perturbation theory \cite{roberge86:_gauge_qcd},
Roberge and Weiss showed that this phase transition is first-order. 
This was confirmed in subsequent  lattice simulations 
\cite{forcrand02:_qcd}. While it was expected that the RW transition is
a signature of the deconfined phase \cite{weiss87:_how},  the transition
line, which is parallel to the temperature axis at $\mu_I/T = \pi/3$, terminates at a
temperature above the deconfinement
transition temperature at vanishing chemical potential. 
Since the characteristics of
the endpoint and its implications for the phase diagram at real $\mu$ are still debated \cite{forcrand:_const_qcd,wu07:_phase_qcd_wilson,kouno09:_rober_weiss,d'elia09:_order_rober_weiss_qcd,sakai:_10063408,bonati:_rober_weiss_endpoin_in_n_f_qcd,Aarts}, 
it is interesting to explore 
the phase structure at imaginary chemical
potential in an effective model. The aim of this paper is to
characterize the phase structure especially of the
confinement-deconfinement transition of QCD at imaginary chemical potential 
in the framework of an effective model which exhibits the relevant symmetries. 

In this work we use the 
Polyakov-loop-extended Nambu-Jona-Lasinio (PNJL) model
\cite{fukushima04:_chiral_polyak,ratti06:_phases_qcd}.
  The NJL model \cite{nambu61:_NJLI,nambu61:_NJLII} 
describes many aspects of QCD
related to chiral symmetry \cite{hatsuda94:_qcd_lagran}. This model,
however, lacks confinement. On the other hand, thermal
models with internal gauge symmetry have been studied. These models reveal 
a RW transition \cite{elze87:_gauge,miller88:_therm_abelian}  while chiral symmetry
is not realized. The PNJL model is an effective model of QCD, which 
ameliorates some of the shortcomings of the NJL model by
introducing a coupling of the quark field to
a uniform background gauge field $A_0$. It has been demonstrated that
the PNJL model reproduces  the RW transition
\cite{sakai08:_polyak_nambu_jona_lasin}.
The authors of Ref.~\cite{sakai08:_polyak_nambu_jona_lasin} have 
studied the phase structure of the PNJL model in detail (see
also~\cite{sakai08:_phase_z,sakai08:_vector_qcd,kouno09:_rober_weiss,sakai09:_deter_qcd,sakai:_09083088,sakai:_10063408,kashiwa11:_nonloc_pnjl_model_and_imagin_chemic_poten}). 
These studies indicate that various improvements are necessary in order
to reproduce the lattice data. 
In this paper, applying the simplest
interaction term in the PNJL model as introduced in Ref.~\cite{sakai08:_polyak_nambu_jona_lasin},
we focus on differences in the behaviour of the order parameters dependently
on the parametrizations of the effective
Polyakov loop potential. 
We characterize the phase structure qualitatively
through a systematic comparison of the results for different Polyakov
loop potentials and give perspectives on the nature of the phase
transitions at imaginary chemical potential.

 We introduce a new quantity, which characterizes the
 confinement-deconfinement transition based on the characteristic
 dependence  of the chiral condensate on  the imaginary chemical
 potential. The relation of this parameter to the so called
dual order parameter~\cite{bilgici08:_dual_polyak} is discussed.

The paper is organized as follows: in the
next section we briefly review the basic properties
of the QCD partition function which are relevant for this study. 
The model is introduced in Sec.~\ref{sec:model} and results
of the numerical calculation are presented in Sec.~\ref{sec:results}.
In Sec.~\ref{sec:dual}, we discuss the parameters characterizing
the confinement-deconfinement transition and finally in
Section \ref{sec:summary} we summarize. 

\section{General properties of the QCD partition function at imaginary
 chemical potential}

\begin{figure}[!t]
 \includegraphics[width=3.375in]{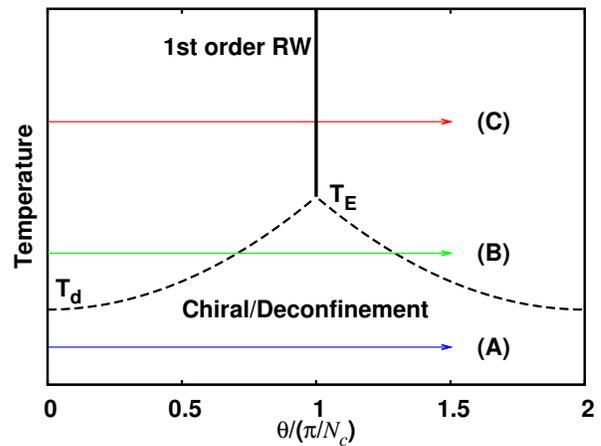}
 \caption{Schematic phase diagram on $T-\theta$ plane. The solid line
 denotes the first-order Roberge-Weiss transition line which terminates
 at $T=T_E$, while the dashed line shows the transition line of the chiral and
 confinement-deconfinement transitions. The arrows labeled A--C indicate
 paths probed in Sec.~\ref{sec:results}. }
 \label{fig:schematic_pd}
\end{figure}

The partition function of the
$SU(N_c)$ gauge theory with fermions, is characterized  by the number operator
$\hat{N}=\int d^3 x \, q^\dagger q$, at imaginary  chemical potential 
$\theta=\mu_I/T = \beta \mu_I$.
\begin{equation}
 Z_G(T,V,\theta) = \text{Tr}\left[ e^{-\beta H + i\theta \hat{N}} \right].
\end{equation}

 The partition function can be expressed in terms of the functional integral 
\begin{equation}\label{part-func}
 Z_G(T,V.\theta)  = \int \mathcal{D}q\mathcal{D}\bar{q}\mathcal{D}A_\mu
e^{-S_E(\beta,\theta)}
\end{equation}
where $S_E$ is the Euclidean action
\begin{equation}
 S_E=\int_{0}^{\beta}d\tau \int d^3 x \, \bar{q}(\gamma\cdot D-m)q
  -\frac{1}{4}G^2 -i \frac{\theta}{\beta}q^\dagger q.
\end{equation}
In (\ref{part-func}), the gauge and quark field obey
periodic and anti-periodic boundary conditions in the temporal
interval $[0,\beta]$, respectively. 
The particle-antiparticle
symmetry implies, 
that $Z_G$ is an even function of $\theta$,
$Z_G(T,V,-\theta) = Z_G(T,V,\theta)$.

By performing the change of variables $q\rightarrow e^{i\theta \tau/\beta}q$, 
the explicit dependence imaginary chemical potential in the action can
be removed and converted into a modified boundary condition
 \cite{weiss87:_how,roberge86:_gauge_qcd}
\begin{equation}
 q(x,\beta) = -e^{i\theta}q(x,0)\label{eq:boundary_immu}.
\end{equation}
Then $Z(N_c)$ transformation
\begin{equation}
 q \rightarrow Uq, \quad A_\mu \rightarrow UA_\mu U^{-1} -(i/g)
  (\partial_\mu U)U^{-1}
\end{equation}
with $U(x,\beta) = \exp(2\pi i k/N_c)U(x,0)$ leaves the action 
and the functional measure invariant,
but modifies the boundary  condition  
\begin{equation}
 q(x,\beta) = -e^{i\theta}e^{\frac{2\pi i}{N_c}k}q(x,0),\label{eq:boundary_immu_2}
\end{equation}
where $k$ is  an integer \cite{weiss87:_how,roberge86:_gauge_qcd}.
A comparison of Eqs. (\ref{eq:boundary_immu_2}) and (\ref{eq:boundary_immu})
reveals that the partition function is  periodic with respect to finite shifts of 
$\theta$
\begin{equation}\label{eq:part-func-periodic}
 Z_G(T,V,\theta) = Z_G\left(T,V,\theta+ \frac{2\pi k}{N_c}\right).
\end{equation}
This periodicity is a remnant of the $Z(N_c)$ symmetry, the center symmetry of the $SU(N_c)$
gauge group. In the presence of fermions the $Z(N_c)$ symmetry is explicitly broken.
In other words, the partition function is invariant under the
transformation $\theta \rightarrow \theta+2\pi k/N_c$ combined with
the $Z(N_c)$ transformation. This symmetry was dubbed ``extended $Z(N_c)$
symmetry'' in Ref.~\cite{sakai08:_polyak_nambu_jona_lasin}.
It is easy to see that  the thermodynamic potential $\Omega=-T\ln Z_G$
and the chiral condensate 
$\sigma\equiv \langle\bar{q}q\rangle = -\frac{1}{V}\frac{\partial
\Omega}{\partial m_0}$, 
with $m_0$ being the current quark mass,
have the same periodicity. 

Roberge and Weiss noted the existence of the first-order
transition at $\theta = \pi/N_c +2\pi k/N_c$ in the deconfined phase
~\cite{roberge86:_gauge_qcd}. 
It was expected that the RW transition takes place at the same temperature
as the confinement-deconfinement transition temperature $T_d$. 
Lattice simulations, however, showed that the endpoint of the RW transition 
is at a temperature $T_E$, which is  higher than $T_d$. 
A schematic phase diagram for imaginary chemical potantial is 
shown in Fig.~\ref{fig:schematic_pd}. The nature of the chiral and
confinement-deconfinement transition at finite $\theta$ is not fully
understood yet. First, there is no a priori reason that these two transitions
coincide. LGT simulations, however, show that the two transitions
take place at approximately the same temperature
\cite{forcrand02:_qcd,d'elia03:_finit_qcd,wu07:_phase_qcd_wilson}. 
Second, the order of the transition depends on the quark mass and the
number of flavors. For $N_f=2$ and $N_c=3$ (this case will be explored in this paper)
Ref.~\cite{wu07:_phase_qcd_wilson} shows that for
$m_\pi/m_\rho \simeq 0.9$ the chiral and confinement-deconfinement
transition at $\theta=0.92(\pi/3)$ is of the crossover type, implying
that the transition at the RW endpoint is second order.
On the other hand,
Refs.~\cite{d'elia09:_order_rober_weiss_qcd,forcrand:_const_qcd} have demonstrated that for light  and heavy
quark masses, the phase transition at $T_E$ is first-order. 
In this case, the first-order RW line continues along  the
dashed lines and  terminates at a critical endpoint which is located at a 
value of the imaginary chemical potential
 $0<\theta<\pi/N_{c}$. 

In this paper, we do not attempt to obtain a fit of the PNJL model to lattice
results at imaginary chemical potential. Rather,  we focus on exploring qualitative 
features of the phase structure of the model and their origin.

\section{PNJL model}
\label{sec:model}

\subsection{Formulation} 
Studies of the phase diagram of strongly at imaginary chemical potential within
effective models, requires a model with the same symmetry
structure as QCD. The most important symmetries which must be accounted for 
are the extended $Z(3)$ 
and the chiral symmetries. In this article we employ the Polyakov loop extended
Nambu-Jona-Lasinio (PNJL) model with $N_f=2$ and $N_c=3$
\cite{fukushima04:_chiral_polyak,ratti06:_phases_qcd}, which respects the above mentioned symmetries.  

The Lagrangian of the two-flavor PNJL model with a four-quark interaction is given by
\begin{gather}
 \mathcal{L} = \bar{q} (i\gamma_\mu D^\mu -m_0)q + G_s [ (\bar{q}q)^2 +
  (\bar{q}i \gamma_5 \vec{\tau}q)^2] \nonumber\\
 - \mathcal{U}(\Phi[A],\Phi^*[A];T).\label{eq:PNJL_lagrangian}
\end{gather}
In the covariant derivative $D^\mu = \partial^\mu - i A^\mu$, only the temporal 
component $A_0$ of the gluon field $A_\mu = gA_\mu^a\lambda^a/2$ is included.
The gluon field is treated as a classical background field, whose dynamics 
is encoded in the effective potential $\mathcal{U}$. The gluon
field is expressed in terms of the traced Polyakov loop and its conjugate
\begin{equation}
 \Phi = \frac{1}{3}\langle\text{Tr}_cL \rangle,\quad \Phi^* = \frac{1}{3}\langle\text{Tr}_cL^\dagger\rangle;
\end{equation}
where the trace is taken over color space and

\begin{equation}
 L(\boldsymbol{x}) = \mathcal{P}\exp\left[ i\int_{0}^\beta d\tau A_4(\boldsymbol{x},\tau)\right]. 
\end{equation}
Here   $A_4 = iA_0$ and and $\mathcal{P}$ denotes the path ordering in Euclidean time $\tau$.
In the Polyakov gauge the matrix $L$ reduces to the diagonal form 
$L=\text{diag}(e^{i\phi_1},e^{i\phi_2}, e^{-i(\phi_1+\phi_2)})$
\cite{fukushima04:_chiral_polyak}. Note that in general $\Phi^*$ is not the complex
conjugate of $\Phi$. At real
chemical potential both $\Phi$ and
$\Phi^*$ are real and at $\mu\neq 0$ their  values differ \cite{sasaki07:_suscep_polyakov}. At imaginary
chemical potential, $\Phi^*$ is the complex conjugate of $\Phi$ and they
have non-zero imaginary parts
\cite{sakai08:_polyak_nambu_jona_lasin}. Therefore, in the discussion below, we use the notation
\begin{align}
 \Phi&= |\Phi| e^{i\phi}, \\
 \Phi^* &= |\Phi| e^{-i\phi}
\end{align} 
for imaginary $\mu$. The extended $Z(3)$
symmetry leads to the following properties of the Polyakov loop \cite{sakai08:_polyak_nambu_jona_lasin}
\begin{align}
 \left|\Phi\left(\theta+\frac{2\pi k}{3}\right)\right| &= |\Phi(\theta)|,\\
 |\Phi(-\theta)| &= |\Phi(\theta)|,\\
 \phi\left(\theta+\frac{2\pi k}{3}\right)&= \phi(\theta) - \frac{2\pi k}{3}\label{eq:Poly_Phase_sym},\\
 \phi(-\theta)&= -\phi(\theta).
\end{align}

In the mean-field approximation
the thermodynamic potential  can be written in terms of quark quasiparticles
with a dynamical mass $M$, momentum $p=|\boldsymbol{p}|$, and energy $E_p=\sqrt{p^2+M^2}$
\cite{fukushima04:_chiral_polyak,sakai08:_polyak_nambu_jona_lasin} 
\begin{align}
 \Omega(T,V,\theta)& = -4 V \int \frac{d^3p}{(2\pi)^3}
  \left[ 3(E_p - E_p^0) \right.\nonumber\\
  & \left. + \frac{1}{\beta}\ln[1 + 3(\Phi+\Phi^* e^{-\beta
   E^-_p})e^{-\beta E^-_p} + e^{-3\beta E^-_p}] \right. \nonumber\label{eq:thermo_pot} \\
 & \left.+ \frac{1}{\beta}\ln[1 + 3(\Phi^*+\Phi e^{-\beta
   E^+_p})e^{-\beta E^+_p} + e^{-3\beta E^+_p}]
   \right] \nonumber \\
  & + (G_s\sigma^2+\mathcal{U})V.
\end{align}
The term in the first line represents the ultraviolet divergent vacuum fluctuations.
As usual, we introduce a three-momentum cutoff $\Lambda$ in the
momentum integration to regularize the vacuum contribution. 
We subtract the vacuum term with the single-quark energy
$E_p^0 = \sqrt{p^2+m_0^2}$, so that the vacuum contribution vanishes when the
chiral symmetry is restored, as in Ref.~\cite{sasaki07:_suscep_polyakov}.
We use a short-hand notation, where the imaginary chemical potential is subsumed in $E^\pm_p =
E_p\pm i\theta/\beta$. The dynamical mass $M$ is
related to the current quark mass and the chiral condensate
$\sigma=\langle \bar{q}q \rangle$ by
$M=m_0-2G_s \sigma$. The term $G_s\sigma^2$ in the last line is due to the 
meson potential in the Lagrangian (\ref{eq:PNJL_lagrangian}).

\begin{table}[!t]
 \caption{Parameters in the polynomial potential \eqref{eq:pot_poly}.}
 \label{tbl:poly_pot}
 \begin{ruledtabular}
  \begin{tabular}{ccccccc}
  $T_0$[MeV]&$a_0$&$a_1$&$a_2$&$a_3$&$b_3$&$b_4$ \\ \hline
  $270$&$6.75$&$-1.95$&$2.625$&$-7.44$&$0.75$&$7.5$ \\
  \end{tabular}
 \end{ruledtabular}

 \caption{Parameters in the logarithmic potential \eqref{pot_log}.}
 \label{tbl:log_pot}
 \begin{ruledtabular}
  \begin{tabular}[t]{ccccc}
   $T_0$[MeV]&$a_0$&$a_1$&$a_2$&$b_3$ \\ \hline
   270&3.51&$-2.47$&$15.22$&$-1.75$ \\
  \end{tabular}
 \end{ruledtabular}
\end{table} 

So far two types of the Polyakov loop effective potential $\mathcal{U}$ have
been widely used. The polynomial potental has as a general $Z(3)$ symmetric form  
\cite{pisarski00:_quark_z_wilson,ratti06:_phases_qcd}:
\begin{equation}
 \frac{\mathcal{U}_{\text{poly}}}{T^4} = -\frac{b_2(T)}{2}\Phi^* \Phi -
  \frac{b_3}{6}
	\left[\Phi^3 + (\Phi^*)^3\right]+\frac{b_4}{4}(\Phi^* \Phi)^2\label{eq:pot_poly}
\end{equation}
with 
\begin{equation}
 b_2(T) =  a_0 + a_1\frac{T_0}{T} + a_2\left( \frac{T_0}{T} \right)^2 +
  a_3 \left( \frac{T_0}{T} \right)^3.
\end{equation}
The coefficients are determined by fitting the equation of
state and the expectation value of the Polyakov loop to lattice data
of pure gauge theory 
\cite{Boyd_NPB469,kaczmarek02:_heavy_quark_antiq_free_energ} in
Ref.~\cite{ratti06:_phases_qcd}.
In the other widely used variant, a logarithmic potential motivated by
the strong coupling expansion is implemented~\cite{fukushima04:_chiral_polyak,roessner07:_polyak}
\begin{align}
 \frac{\mathcal{U}_{\text{log}}}{T^4} = &-\frac{a(T)}{2}\Phi^*
 \Phi+b(T) \nonumber\\
 &\times\log\left\{1-6\Phi^* \Phi +4\left[\Phi^3 + (\Phi^*)^3\right] -3(\Phi^* \Phi)^2\right\}\label{pot_log}
\end{align}
where
\begin{equation}
 a(T)= a_0 + a_1\frac{T_0}{T} + a_2 \left( \frac{T_0}{T}
						   \right)^2,\quad 
 b(T) = b_3\left(\frac{T_0}{T}\right)^{3}.
\end{equation}
The above parameterization for the temperature dependency was
introduced in \cite{roessner07:_polyak} and the constants are determined
by fitting lattice data of pure SU(3) theory. 
In Ref.~\cite{fukushima04:_chiral_polyak}, a similar functional form in
$\Phi$ but different parameterization of the temperature dependence was
introduced. In this potential, however, one of the parameters is
fixed to reproduce the simultaneous crossover transition for chiral and
deconfinement transition rather than the pure gauge theory except for the
transition temperature $T_0\simeq 270$ MeV. 
We refer to \cite{Fukushima2008} for discussion.
This difference makes it difficult to perform a systematic
comparison of effect of quarks near the deconfinement transition.
If we re-fit the parameters to reproduce the pure SU(3) lattice data,
we expect to have similar results to those from the logarithmic
potential \eqref{pot_log} since the target space and the transition temperature are
the same. 
In this paper, we focus on the two potentials
Eqs.~\eqref{eq:pot_poly} and \eqref{pot_log} which equally reproduce the
Polyakov loop and thermodynamics as well as the first order
confinement-deconfinement phase transition.
We use the parameters determined in Refs.~\cite{ratti06:_phases_qcd} and
\cite{roessner07:_polyak}. For convenience, they are summarized in Tables
\ref{tbl:poly_pot} and \ref{tbl:log_pot}. 

The order parameters, chiral condensate $\sigma$ (or
dynamical mass $M$), modulus of the Polyakov loop $|\Phi|$, and the phase
of the Polyakov loop $\phi$ are determined numerically by solving
the coupled equations of motion
\begin{equation}
 \frac{\partial \Omega(T,V,\theta;M,|\Phi|,\phi)}{\partial X_i} = 0, \quad X_i = M, |\Phi|, \phi .\label{eq:gapeqfull}
\end{equation}
The phase diagram in the $T-\theta$ plane 
of the polynomial potential model 
\eqref{eq:pot_poly} has been studied in
Refs.~\cite{sakai08:_polyak_nambu_jona_lasin,sakai08:_phase_z}.
In this model, the first-order Roberge-Weiss transition at 
$\theta=\pi/3\pm 2\pi k/3$, the 
second-order chiral transition in the chiral limit and the crossover
 one at finite quark mass as well as the crossover confinement-deconfinement
 transition were found. 
However, these features,  depend on the choice of the Polyakov loop
effective potential and further quark interaction terms are required to reproduce lattice results
quantitatively~\cite{sakai:_10063408,sakai08:_phase_z}. 

In this paper, we restrict ourselves to the simplest quark-quark interaction
form, as shown in Eq.~\eqref{eq:PNJL_lagrangian} and focus on behavior of the
order parameters in the $T-\theta$  plane for the  polynomial and
logarithmic  Polyakov loop potentials. 

\subsection{Some analytic insights}
\label{sec:analyticinsight}

Before proceeding to the full
numerical computation, it is useful to explore the general properties
of the thermodynamic potential \eqref{eq:thermo_pot} analytically in a few limiting cases.
The momentum integration in Eq.~\eqref{eq:thermo_pot} can be carried out analytically if we first
expand the logarithmic terms in the integrand in powers of $e^{-\beta E_p} \ll 1$. 
We thus find, keeping terms up to order $(e^{-\beta E_p})^3$,
\begin{align}
 \Omega \simeq &(G_s\sigma^2+\mathcal{U})V - \Omega_0 -\frac{2V}{\beta
 \pi^2} \nonumber \\
 \times&\left[ 3(\Phi e^{i\theta} + \Phi^* e^{-i\theta})\intop_0^\Lambda dp p^2
 e^{-\beta E_p}\right. \nonumber \\
 +&\frac{3}{2}\left\{ e^{2i\theta}(2\Phi^* - 3\Phi^2) +
 e^{-2i\theta} ( 2 \Phi - 3\Phi^{*2}) \right\} \intop_{0}^{\Lambda} dp p^2
 e^{-2\beta E_p} \nonumber \\
 +&\left\{ 2(1-9\Phi \Phi^*)\cos 3\theta + 9 e^{3i\theta}\Phi^3 + 9
 e^{-3i\theta}\Phi^{*3} \right\} \nonumber\\
 \times& \left.\intop_{0}^{\Lambda} dp p^2 e^{-3\beta E_p} \right].\label{eq:omega_approx}
\end{align}
Here, $\Omega_0$ is the temperature independent vacuum term. 
While it  is necessary to introduce a finite cutoff for this term, due to the
non-renormalizability of the PNJL model, taking $\Lambda\rightarrow
\infty$ in the thermal part does not affect the qualitative features
discussed below. Thus the momentum integration of the thermal part can be
carried explcitily resulting in  modified
Bessel functions $K_n$.

We first consider the low temperature limit $\Phi=\Phi^*=0$ in order to explore
the effect of the
Polyakov loop in the confined phase. In this case the Polyakov loop effective potential,
$\mathcal{U}$, vanishes. Furthermore, the gap equation for the dynamical mass $M$, 
obtained from Eq.~\eqref{eq:omega_approx}, $\partial \Omega/\partial M = 0$, reduces to
\begin{align}
 M \simeq m_0&+\frac{6G_s}{\pi^2}[M^3f(\Lambda/M)-m_0^3f(\Lambda/m_0)]
 \nonumber \\
& -\frac{8G_s M^2 T \cos 3\theta}{\pi^2}K_1(3M/T),\label{eq:gap_confine}
\end{align}
where
\begin{equation}
 f(x) = x\sqrt{x^2+1}-\ln (x+\sqrt{x^2+1}).
\end{equation}
Note that, for $3M/T \gg 1$, 
\begin{equation}
 K_1(3M/T)\sim \sqrt{\frac{\pi T}{6M}}e^{-3M/T}.
\end{equation}

The gap equation (\ref{eq:gap_confine}) implies that the $\theta$ dependence of the dynamical
mass is completely determined by $\cos3\theta$ term. Consequently, $M$ is a periodic function of
$\theta$ with the period $2\pi/3$, as expected. 
For small temperatures, quark degrees of freedom are strongly suppressed, $\sim e^{-3M/T}$, since, for a vanishing 
Polyakov loop, only three-quark clusters survive. 
The chiral phase transition takes place when the thermal
contribution in the gap equation is of the same order as the vacuum term. If the condition $\Phi=\Phi^*=0$ is strictly enforced, the chiral transition is shifted to very high temperatures. Because the thermal excitation of quarks in this limit is possible only in three quark clusters, the resulting thermodynamics is qualitatively similar to that of the nucleonic NJL model, which also yields a very high chiral transition temperature \cite{sasaki10:_therm_of_dense_hadron_matter}.

\begin{figure}[!t]
 \includegraphics[width=3.375in]{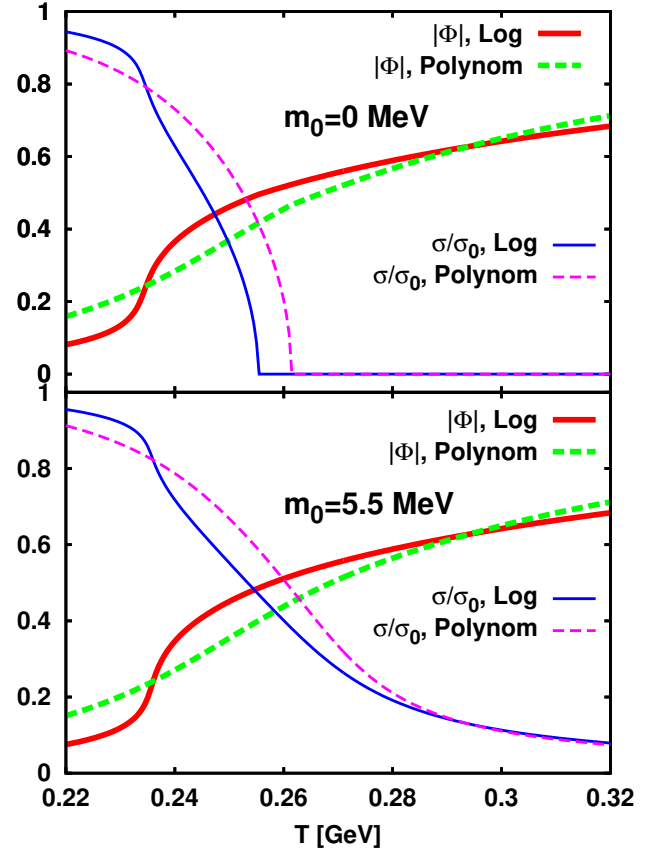}
 \caption{Temperature dependence of the chiral condensate normalized by
 the value at $T=0$ and the Polyakov loop at vanishing chemical
 potential. Upper panel shows the result in the chiral limit while the
 lower stands for the physical quark mass.}
 \label{fig:mu0}
\end{figure}

In the (naive) high-temperature limit, $\Phi=\Phi^*\to1$, the gap equation
reduces to that of the ordinary NJL model,
\begin{align}
 M =m_0&+\frac{6G_s}{\pi^2}[M^3f(\Lambda/M)-m_0^3f(\Lambda/m_0)]
 \nonumber \\
 &-\frac{24G_s M^2 T \cos \theta}{\pi^2}K_1(M/T).\label{eq:gapeq_NJL_boltzmann}
\end{align}
Now, the $\theta$ dependence of the dynamical mass is determined
by $\cos\theta$ and the thermal factor is proportional to $K_1(M/T)$,
appropriate for the thermal excitation of single quarks.
In this case, the dynamical mass
has lost the original periodicity of the partition function (\ref{eq:part-func-periodic}),
which is respected in the low-temperature limit \eqref{eq:gap_confine}.

At $\theta=\pi/2$ the thermal contribution in (\ref{eq:gapeq_NJL_boltzmann}) vanishes and consequently the 
dynamical mass equals its vacu\-um value, irrespective of
temperature. Hence, in this approximation the phase boundary is shifted to higher temperatures as the imaginary 
chemical potential is increased, and eventually approaches $T=\infty$ in the limit $\theta \to \pi/2$. In general, the 
thermal contribution is of the form
$-\sum_{n}L_n \cos n\theta K_1(nM/T)$ with $L_n > 0$. Due to the higher order terms, the transition temperature remains finite, but the positive curvature of the phase boundary persists~\footnote{Owing to the reflection
symmetry and periodicity, it is sufficient to consider the interval $0 \leq \theta
\leq \pi/3$.}. On the other hand, for real 
chemical potentials the leading thermal contribution is proportional to $T \cosh\beta\mu$. Since $\cosh x$ is an increasing function of $x$, this implies that
the chiral transition temperature decreases as the real chemical potential grows.

In the high-temperature limit, as implemented above, the original periodicity of the partition function is lost because   
the phase of the Polyakov loop is neglected.
In fact, at imaginary chemical potential, the high temperature limit of the PNJL model
is in general not the NJL model. 
The Roberge-Weiss transition, characterized by
discontinuous jumps of the phase $\phi$,  preserves the
periodicity $2\pi/3$ in the deconfinement phase.

\section{Behavior of the order parameters at imaginary chemical potential}
\label{sec:results}
\begin{figure}[!t]
 \includegraphics[width=3.375in]{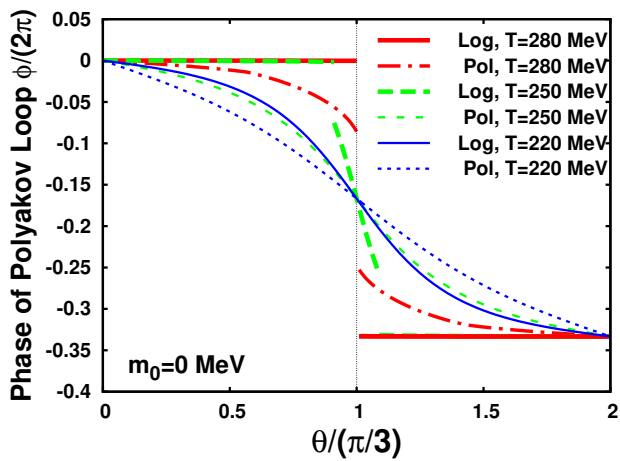}
 \caption{Phase of Polyakov loop as a function of $\theta$ for various
 temperatures in the chiral limit. As in Fig.~\ref{fig:mu0}, each line
 stands for the results of the logarithmic Polyakov loop
 potential and the polynomial one at the different temperatures, respectively. }
 \label{fig:phi-theta_m0}
\end{figure}

We now discuss the characteristics of the order parameters obtained by solving the full gap equation
Eq.~\eqref{eq:gapeqfull}. Besides the Polyakov loop potentials, given in
the previous section, the model has three parameters in the
fermion sector. We use
\begin{align}
 G_s &= 5.498 \text{GeV}^{-2},\label{eq:coupling_physical}\\
 \Lambda &= 0.6315 \text{GeV},
\end{align}
which reproduce the vacuum pion mass and pion decay constant
at zero temperature and density, when the current quark mass is fixed to
the value $m_0=5.5$ MeV ~\cite{hatsuda94:_qcd_lagran}.

In what follows, we compare the results obtained 
for a finite pion mass with those corresponding to the chiral limit, $m_0=0$.
In the latter case, the model belongs to 
the universality class of the three dimensional $O(4)$ spin model
and exhibits a second-order phase transition at finite temperature 
and small values of the real chemical potential.

With the parameter set given above, we find the chiral condensate and the
Polyakov loop  shown in Fig.~\ref{fig:mu0}. The dependence of the order parameters on the temperature shows that in 
the chiral limit, the chiral transition is indeed second-order, while the confinement-deconfinement transition is of the 
cross over type. For finite quark mass, $m_0=5.5$ MeV, shown in the lower panel, the chiral order parameter 
and the Polyakov loop both exhibit smooth crossover transitions. Thus, the explicit symmetry breaking induces a 
qualitative change of the chiral condensate, while for the Polyakov loop this dependence 
is negligible. Furthermore, a comparison of  the
two parametrizations of the Polyakov loop effective potential shows that the transition is smoother for the polynomial
potential than for the logarithmic one. 

\begin{figure}[!t]
 \includegraphics[width=3.375in]{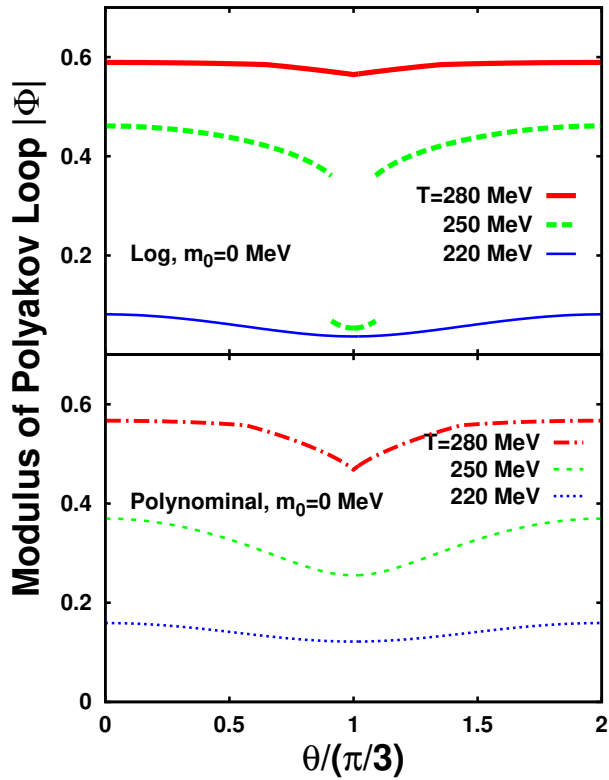}
 \caption{Modulus of Polyakov loop as a function of $\theta$ in the
 chiral limit. Upper panel: Logarithmic potential, Lower panel: Polynomial
 potential. The temperatures are the same as
in Fig.~\ref{fig:phi-theta_m0}.}
 \label{fig:poly-theta_m0}
\end{figure}


\subsection{Behavior across $\theta=\pi/3$}
\label{subsec:thetadep}
We now consider the order parameter as a function of $\theta$
close to  $\theta=\pi/3$.
In Figure \ref{fig:phi-theta_m0} we show the phase
of the Polyakov loop $\phi$ as a function of $\theta$ in the chiral
limit. We do not show results for non-zero quark mass, since the results are indistinguishable from 
those shown in Fig.~\ref{fig:phi-theta_m0}.  

 At low temperatures (arrow A in Fig.~\ref{fig:schematic_pd}), the
phase of the Polyakov loop changes smoothly from 0 at $\theta=0$ to 
$-\pi/3$ at $\theta=\pi/3$. Subsequently, the phase continues to decrease and
finally approaches to $-2\pi/3$ at $\theta=2\pi/3$, as required by the
symmetry, Eq.~\eqref{eq:Poly_Phase_sym}.
This behavior is independent of the choice of $\mathcal{U}$.
At temperatures beyond $T_{E}$, the transition is first order (arrow C in Fig.~\ref{fig:schematic_pd}). 
For instance at $T = 280$ MeV, 
the phase jumps from $\phi=0$ to $-2\pi/3$ at $\theta=\pi/3$. This is the Roberge-Weiss transition
 \cite{roberge86:_gauge_qcd}, where the phase of the Polyakov loop jumps from one
 $Z(3)$ sector to another. The RW transition is common to both parametrizations of
$\mathcal{U}$. This is natural, since the RW transition is a consequence of
of the $Z(3)$ symmetry of the pure gauge theory, which is incorporated in both potentials.
The detailed behavior around $T_E$ is, however, different between the
two potentials. Thus, at $T=250$ MeV for the logarithmic
potential, which is below the endpoint of the RW transition ($T_{E}\simeq 255$ MeV), the phase is discontinuous at $\theta\neq \pi/3$. This implies that the phase boundary, which is crossed by arrow B in Fig.~\ref{fig:schematic_pd}, is first order at this temperature. By contrast, in the polynomial case the phase is a smooth function of $\theta$ at any temperature below the RW endpoint ($T_{E}\simeq 275$ MeV). As we discuss below,
this reflects the different order of the RW endpoint for the two potentials. 

We note that the logarithmic potential is defined in a limited domain,
characterized by positivity of the argument of the logarithm, while for   
the polynomial potential there is no such restriction. In fact, for high temperatures
(e.g. $T=280$ MeV) the Polyakov loop, plotted for the polynomial potential as a function 
of $\theta$ in the complex $\Phi$ plane, leaves the 
so called target space, defined by requiring that the 
logarithmic potential is real \cite{wozar06:_phase_z_polyak}.

In Figure \ref{fig:poly-theta_m0} we show the modulus of the Polyakov loop
$|\Phi|$ as a function of $\theta$.
At high temperatures, the RW transition is manifested by a cusp at $\theta=\pi/3$, 
while at lower temperatures $|\Phi|$ varies smoothly for both parametrizations of the 
Polyakov-loop potential. However, at intermediate temperatures, 
the two potentials yield qualitatively different results, as illustrated by the discontinuities 
in $|\Phi|$ obtained for the logarithmic potential near $\theta = \pi/3$ at $T=250$ MeV. 
For the polynomial potential we find a continuous confinement-deconfinement transition 
at imaginary chemical potential, while for the logarithmic potential the transition is first order at intermediate 
temperatures.
We return to this point in the following subsection. Here we note only that the
first order transition is reflected also in a sudden change of the
phase at $T=250$ MeV, as shown in Fig.~\ref{fig:phi-theta_m0}. 

\begin{figure}[!t]
 \includegraphics[width=3.375in]{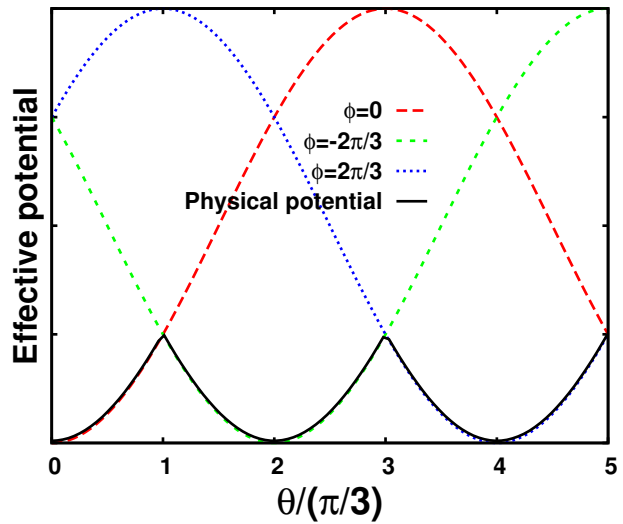}
 \caption{Schematic effective potential for the RW transition.}
 \label{fig:cos-potential}
\end{figure}

Within the PNJL model, the transition from one $Z(3)$ 
sector to another can be understood in the following way.
In the high temperature limit, the dominant contribution to the thermodynamic potential is given by the
single quark excitation term in Eq.~\eqref{eq:omega_approx} (the first
term in the square bracket), which yields a contribution to $\Omega \sim
-\cos(\theta+\phi)$. The only additional $\phi$-dependent term in $\Omega$ 
is the Polyakov loop potential $\mathcal{U}$. At high
temperature, $\mathcal{U}$ has the three local minima at
$\phi=0$ and $\pm 2\pi/3$. For each value of
$\theta$, the physical vacuum is obtained by finding the 
absolute minimum of the two terms. As illustrated in Fig.~\ref{fig:cos-potential}, the physical
vacuum changes from one minimum to the next as $\theta$ crosses $\pi/3+2\pi k/3$.
While both potentials have the
periodicity $\Delta\theta=2\pi$, when $\phi$ is artificially fixed in one $Z(3)$ sector,
the complete thermodynamic potential acquires the periodicity
$\Delta\theta=2\pi/3$ owing to the $\phi$ dependence of $\mathcal{U}$. 

\begin{figure}[!t]
 \includegraphics[width=3.375in]{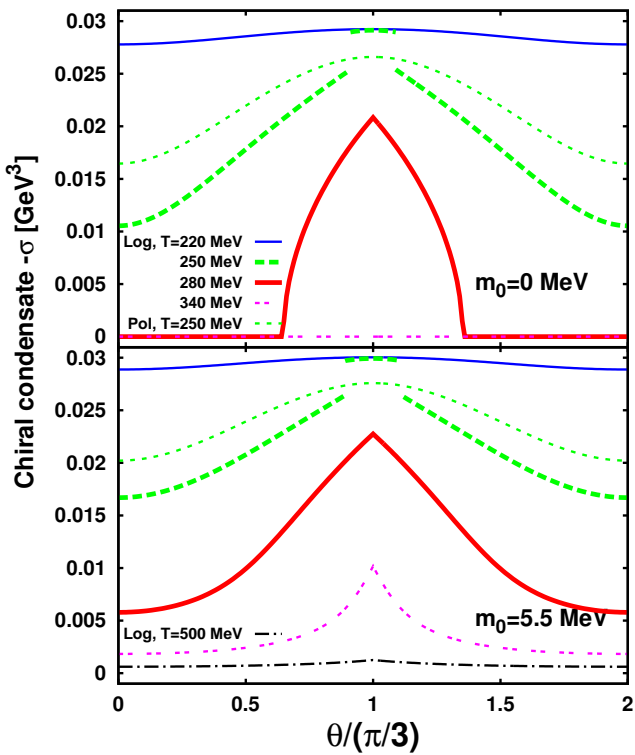}
 \caption{Chiral condensate at various temperatures as a function of
 $\theta$ in the chiral limit. The temperatures associated with the lines are the same as
 Fig.~\ref{fig:phi-theta_m0}. }
 \label{fig:sigma-theta_m0}
\end{figure}
 
Similarly, the chiral condensate $\sigma$ is also continuous  near
$\theta=\pi/3$ at  temperatures lower than $T_E$,
irrespective of the quark mass and the potential, as shown in
Fig.~\ref{fig:sigma-theta_m0}.
At temperatures higher than $T_E$, it develops a cusp for both parametrizations, 
a manifestation of the RW transition. 
Furthermore, at temperatures above $T_\chi(\theta=\pi/3)$,
$\sigma$ vanishes for any value of $\theta$ in the chiral limit, as shown for $T=340$ MeV.
For a finite quark mass, the cusp persists to temperatures much higher than $T_\chi$ but finally
disappears, as shown in Fig.~\ref{fig:sigma-theta_m0} for $m_0=5.5$ MeV and $T=500$ MeV.

In the chiral limit, for temperatures in the interval between $T_{\chi}(\theta=0)$ and $T_{\chi}(\theta=\pi/3)$, there is a second order chiral transition at non-zero $\theta$, as shown in Fig.~\ref{fig:sigma-theta_m0},
(see also arrow B in Fig.~\ref{fig:schematic_pd}). For finite quark masses, this transition is of the cross-over type, 
as illustrated in the lower panel of Fig.~\ref{fig:sigma-theta_m0} for $m_0=5.5$ MeV.
Also the temperature dependence of $\sigma$ is clearly different for the two
potentials. In particular, there is a discontinuity in $\sigma$ near $\theta=\pi/3$
at $T=250$ MeV for the logarithmic potential. The values of  temperature and imaginary chemical potential corresponding to the discontinuity are identical to the ones obtained for the Polyakov loop.

\begin{figure}[!t]
 \includegraphics[width=3.375in]{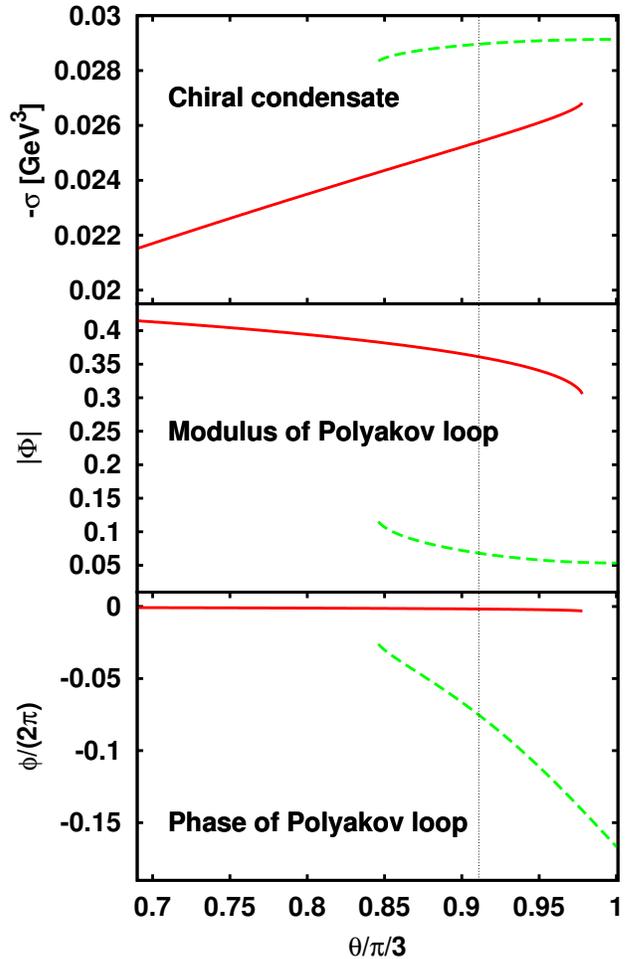}
 \caption{Behavior of the order parameters around a  first order phase
 transition point in the chiral limit. We depict $T=250$ MeV.
 From top to bottom, each panel shows the chiral condensate, modulus of
 the Polyakov loop, and the phase of the Polyakov loop. The vertical line
 (thin dashed) show the critical $\theta$, determined by a Maxwell construction. }
 \label{fig:firstorder}
\end{figure}

\subsection{First-order phase transition at $T < T_{E}$}

In this section we focus on the first order transition found in a limited temperature range below the Roberge-Weiss transition
for the logarithmic potential. In Fig.~\ref{fig:firstorder} we illustrate this result at $T=250$ MeV. In each panel two lines are shown: one is the
solution of the gap equations approaching the transition from small $\theta$, while the
other is obtained by approaching from the opposite side. 

The existence of two solutions in a certain range of $\theta$ shows that there are two local minima in the
thermodynamic potential. The first-order phase transition takes place at the value of $\theta$, where the 
thermodynamic potential in the two local minima is degenerate. At $T=250$ MeV, this happens at $\theta=0.911(\pi/3)$. 
The lines terminate where the corresponding minimum disappears. 
Thus, the system exhibits hysteresis, a characteristic of a first-order phase transition.

Although this transition is related to the confinement-deconfinement transition, the discontinuity 
is reflected also in the chiral condensate $\sigma$, owing to the coupling between the Polyakov loop and 
the chiral order parameter. 

\subsection{The RW endpoint}

\begin{figure}[!t]
 \includegraphics[width=3.375in]{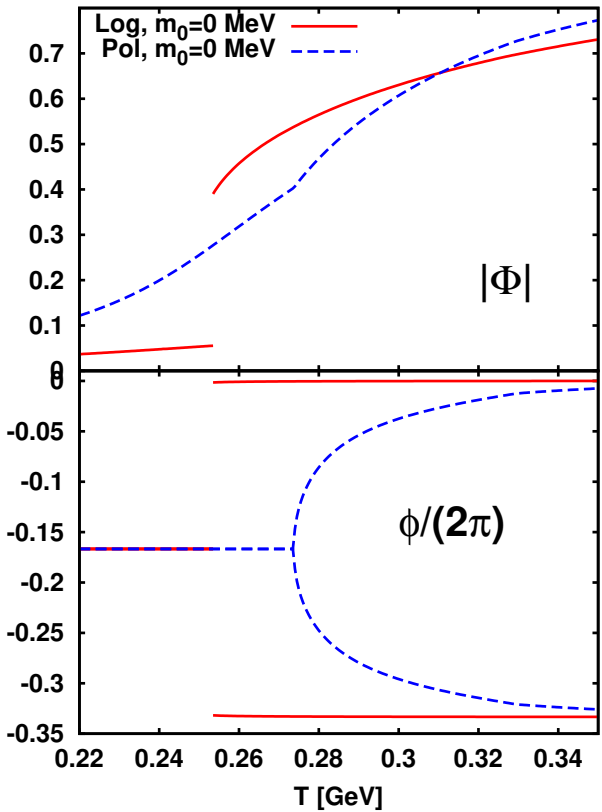}
 \caption{Behavior of $|\Phi|$, and $\phi$ as functions of $T$ at
 $\theta=\pi/3$. }
 \label{fig:RWendpoint}
\end{figure}

The existence of the first-order transition, discussed in the previous section, is closely related to the
characteristics of the RW endpoint. In Fig.~\ref{fig:RWendpoint} we show the temperature dependence of the Polyakov loop
along the line $\theta=\pi/3$ (cf. Fig.~\ref{fig:schematic_pd})
for the two potentials in the chiral limit.
The nature of the RW endpoint, as characterized e.g. by the phase of the
Polyakov loop, differs between the two potentials.
While the polynomial potential yields a continuous transition,
the logarithmic one exhibits a discontinuity in the phase and magnitude of the Polyakov loop. 

Above the RW endpoint ($T=T_{E}$), the phase of the Polyakov loop can take two 
values on the $\theta=\pi/3$ line, corresponding to different $Z(3)$ sectors (cf.~Fig.~\ref{fig:cos-potential}). Thus, in the case of the 
polynomial potential, the phase bifurcates smoothly at the RW endpoint, while for the logarithmic potential, 
the phase changes discontinuously at this point. In the former case, the RW endpoint is a second order point, while in the latter it is a triple
point. We note, however, that a different parameterization for the logarithmic potential
\cite{fukushima04:_chiral_polyak} also yields a second order RW endpoint~\cite{kouno09:_rober_weiss}.

Consequently, the characterization of the RW endpoint depends on the parametrization of the Polyakov loop potential, but, within the framework considered here, it is independent of the value for the quark mass for the logarithmic potential.
On the other hand, in LGT calculations it is found that the nature of the RW endpoint depends crucially on the quark mass; for both two and three flavor QCD it is a second order endpoint at intermediate quark masses and a triple point for large and small masses~\cite{forcrand:_const_qcd,bonati:_rober_weiss_endpoin_in_n_f_qcd}.  
In Ref.~\cite{sakai:_10063408} it is argued that in order to reproduce the quark mass dependence of the RW
transition, a $\Phi$
dependent fermion coupling, motivated by functional renormalization group
analyses \cite{kondo:_towar_qcd,braun_0908.0008}, is required.
This indicates that in QCD the interplay between chiral symmetry
breaking and confinement is more complicated than in the present
model. 

\begin{figure}[t]
 \includegraphics[width=3.375in]{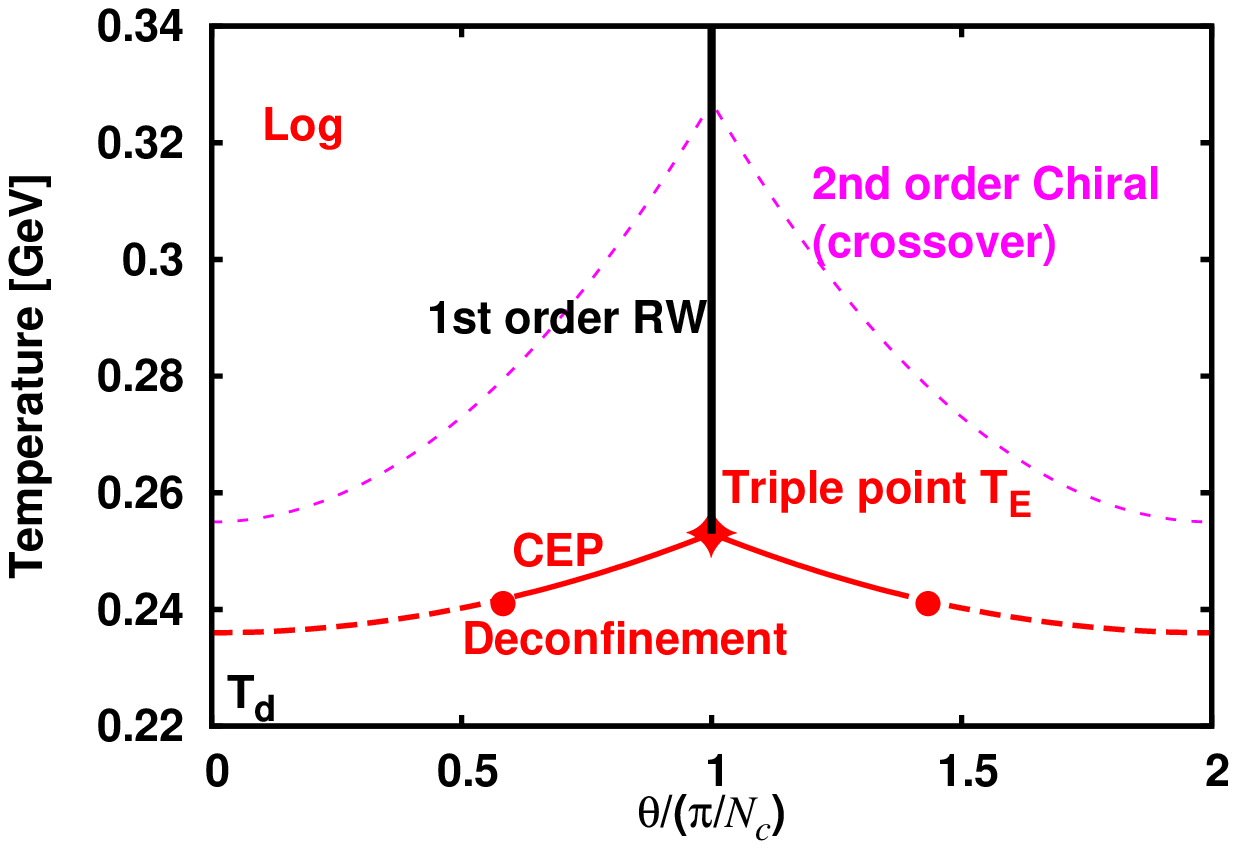}
 \includegraphics[width=3.375in]{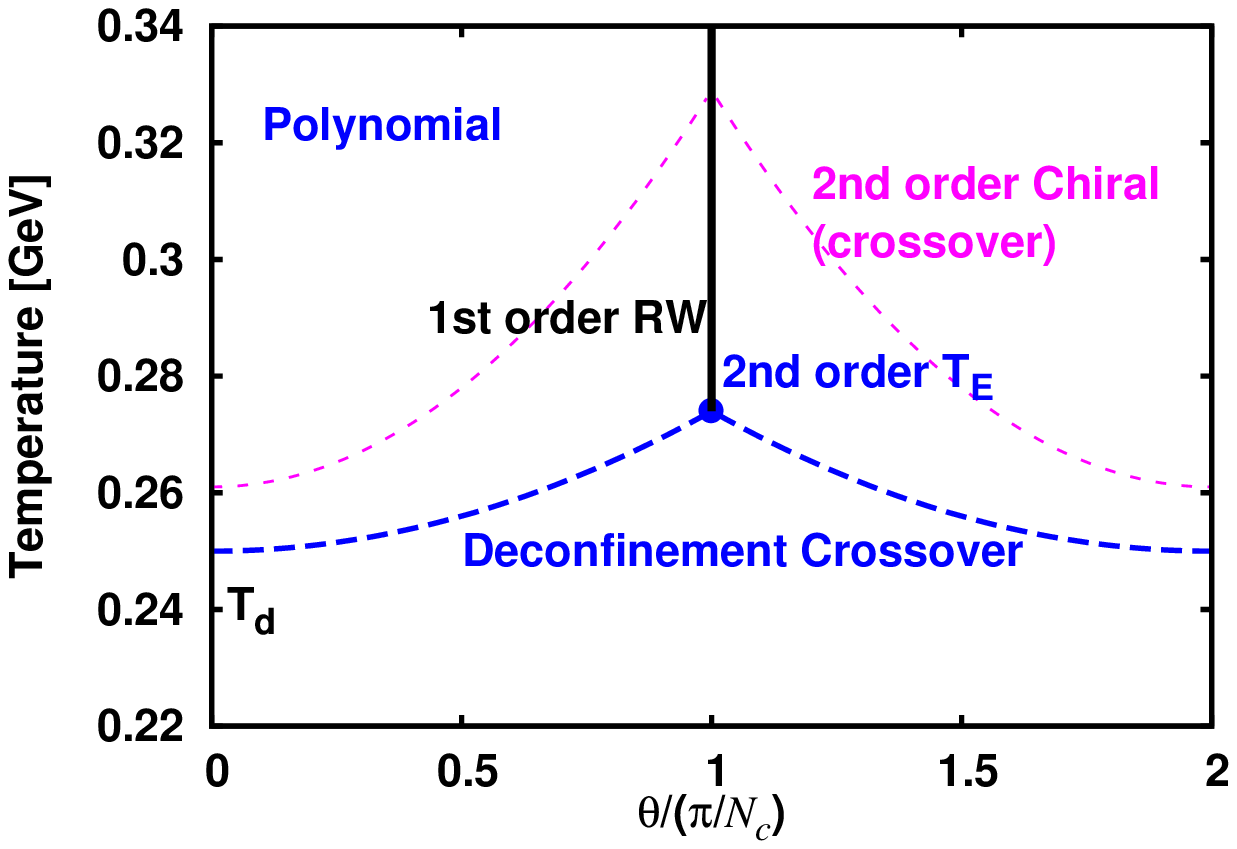}
 \caption{Phase diagram for the logarithmic potential (top) and the
 polynomial potential (bottom).}
\label{fig:9}
 \end{figure}

Finally, in Fig.~\ref{fig:9} we summarize the results on the phase diagram in the $T-\theta$ 
plane for the two Polyakov loop potentials.  The
(pseudo-)critical temperature for the deconfinement transition corresponds to
a maximum of the temperature derivative of the modulus of the Polyakov
loop $d|\Phi|/dT$.
The critical end point (CEP), obtained for
the logarithmic potential at
$0 < \theta < \pi/3$, is a consequence of the triple point at
$T=T_E$  and $\theta=\pi/3$. 

In comparison of the two potentials, 
we have seen that the phase transitions at imaginary chemical
potential are shifted to higher temperatures compared to those at 
real chemical potential. This implies that dynamical quark mass becomes
heavier at fixed temperature (see Fig.~\ref{fig:sigma-theta_m0}) thus
the Polyakov loop potential $\mathcal{U}$, which is independent of
$\theta$, 
makes a dominant contribution to the thermodynamic potential.
Furthermore, at imaginary chemical potential, the target space of the
Polyakov loop is probed through the change of the phase $\phi$.
Therefore, a comparison of the resulting phase diagram at the
imaginary chemical potential region with that obtained in LGT
calculations, yields important constraints on the effective Polyakov
loop potential.

\subsection{Critical endpoint of confinement-deconfinement transition}

In Ref.~\cite{bowman09:_critical} it was found that as the pion mass in
the quark-meson model is reduced from its physical value, 
an additional critical endpoint appears on the phase boundary at small
(real) values of the chemical potential. Since the coupling to the
Polyakov loop is not accounted for in \cite{bowman09:_critical}, the
additional CEP is associated with the chiral phase
transition~\footnote{The fact that the model calculation of
Ref.~\cite{bowman09:_critical} yields a first order chiral transition at
both small and large values of $\mu$ in the chiral limit, is presumably
due to the neglect of the fermion vacuum loop \cite{skokov10:_vacuum_fluct_and_therm_of_chiral_model}.}. In this section we explore the
dependence of the confinement-deconfinement CEP,
which appears at imaginary $\mu$ for the logarithmic potential, on the model parameters. 
We find that the location of this CEP depends on the four fermion coupling constant $G_s$. 
\begin{figure*}[!t]
 \includegraphics[width=6.75in]{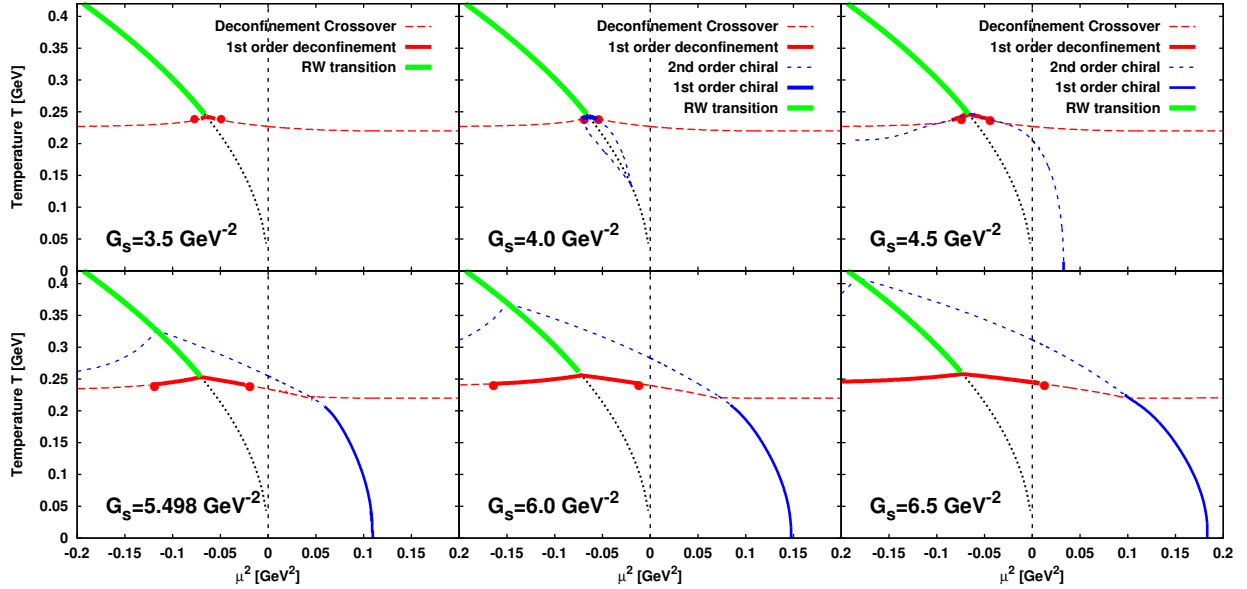}
 \caption{Phase diagrams on $T-\mu^2$ plane for various $G_s$ in the
 chiral limit. Solid
 lines stand for the first order phase transition lines. Dashed lines
 denote the second or crossover transition. For convenience,  $\mu^2=0$
 and $\mu_I/T=\pi/3$ are indicated by the thin and thick dotted lines,
 respectively. The closed circles show the location of the CEP for the
 confinement-deconfinement transition.}
 \label{fig:pd_t-mu2}
\end{figure*}

In Fig.~\ref{fig:pd_t-mu2} we show the phase diagram of the model in the chiral
limit for different values of $G_s$. We include both real and imaginary values of $\mu$ 
by showing the phase boundaries in $T-\mu^2$ plane. The upper
panels show cases where $G_s$ is smaller,
while the lower ones show
cases where it is larger than the reference value \eqref{eq:coupling_physical}.
Lines appearing for $\mu^2 < -(\pi T/3)^2$ (the boundary is indicated by the dotted line)
are images of those in the region $-(\pi T/3)^2 \leq \mu^2 \leq 0 $; the mapping is defined by the
periodicity of the partition function.

We identify the (pseudo)critical temperature by finding the the maximum of the 
derivative of the corresponding order parameter with respect to temperature. For real values of the chemical
potential, the Polyakov loop and its conjugate are real but take on
different values \cite{sasaki07:_suscep_polyakov}. Here we use
$d\Phi/dT$ for the definition of the deconfinement transition. A different 
definition, based e.g. on the Polyakov loop susceptibility
\cite{sasaki07:_suscep_polyakov}, would lead to a slightly different value of the pseudocritical temperature 
of the crossover transition. 

For the first order transition at large $\mu^2$ one finds double
peaks in $d\Phi/dT$ (cf. Fig.~\ref{fig:pd_t-mu2}). Here we identify the position of 
the transition with the maximum which
smoothly extrapolates to the deconfinement transition at vanishing
chemical potential and to the chiral transition at $T=0$, where the peak structure
is simple. We note that any ambiguity in the location of the phase boundary 
does not affect the discussion below.

Qualitatively the features of the phase diagram can be classified as follows.
The NJL sector has a critical coupling
$G_s^{\text{cr}}=\pi^2/(2N_c \Lambda^2)$ for the gap equation to have a
nontrivial solution \cite{hatsuda94:_qcd_lagran}.  This implies that, with the present three-momentum cutoff, 
there is no spontaneous breakdown of chiral symmetry for $G_s < 4.125$ GeV$^{-2}$. Therefore, in the
upper-left panel ($G_s=3.5$ GeV$^{-2}$), the system is everywhere in
the chirally symmetric phase. In this case the RW endpoint is still a
triple point, and the CEP of the deconfinement transition is close by, at $\theta=0.95\pi/3$. 

For $G_s > G_s^{\text{cr}}$, there is a chiral transition at vanishing
chemical potential. As seen in the upper-center panel, there is
a precursor at imaginary chemical potential for
$G_s$ slightly smaller than $G_s^{\text{cr}}$. The chiral symmetry is spontaneously
broken in a small region at intermediate temperatures adjoining the
$\mu^2 = -(\pi T/3)^2$ line. This behavior can be understood along the lines 
presented in Sec.~\ref{sec:analyticinsight}. Although the Boltzmann
approximation, Eq.~\eqref{eq:omega_approx}, might not be a good approximation since the system is in
the chirally symmetric phase even at low temperatures, the Polyakov loop is
small so that the thermal contribution is dominated by the $\cos3\theta$
term as in Eq.~\eqref{eq:gap_confine}. 
Since this term is positive in the $\pi/6 < \theta < \pi/2$, it adds to the vacuum term and a non-trivial
solution appears at finite temperature where the system enters the broken
phase. 

As the temperature is increased further,
however, the Polyakov loop is non-zero and the one- and two-
quark excitations contribute to the gap equation, driving the system 
back into the symmetric phase. Consequently, near the RW endpoint, the chiral and deconfinement transitions occur
simultaneously and the chiral transition is also of first order.
Note that the lower endpoint of the chiral
transition follows the $\mu^2 = -(\pi T/3)^2$ line and arrives at
the origin when $G_s=G_s^{\text{cr}}$. For $G_{s}$ beyond this value, the chiral transition line
enters the $\mu^2 > 0$ half-plane. 

\begin{figure}[!t]
 \includegraphics[width=3.375in]{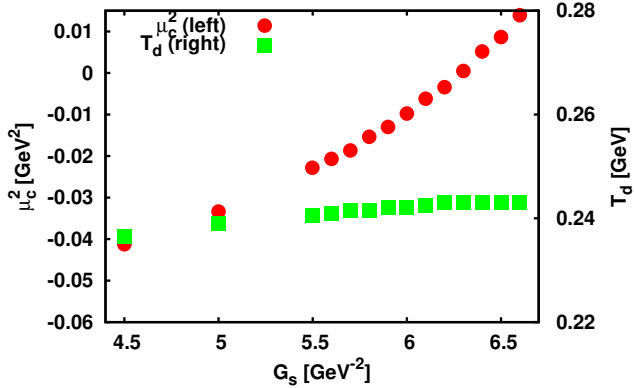}
 \caption{Location of the CEP as a function of $G_s$. The left vertical
 axis shows the chemical potential while the right one denotes the temperature.}
 \label{fig_CEP}
\end{figure}

As $G_s$ is increased beyond $G_s=G_s^{\text{cr}}$, the location 
of the deconfinement CEP moves to larger $\mu^2$ and finally reaches real values of the chemical potential at
$G_s \simeq 6.5$ GeV$^{-2}$. At the same time, the chiral transition line
moves to larger $T$ and $\mu^2$. The behavior of
the CEP can be directly related to changes of the chiral transition with
increasing $G_s$. Increasing $G_s$ leads to a larger dynamical mass 
$M = m_0-2G_s\sigma$ in the
chirally broken phase.  
Moreover, $|\sigma|$ increases with $G_s$ since
a stronger scalar coupling leads to larger quark condensate.
This raises both the dynamical mass and the chiral transition temperature. On the other
hand, a modified $G_s$ does not alter the Polyakov loop sector
of the model. Consequently, near the deconfinement transition, the dynamical mass of the quarks
increase with $G_{s}$ and the system approaches a pure
gauge theory owing to the thermal suppression of quark degrees of freedom. Thus, as shown in
Fig.~\ref{fig_CEP}, the first order phase
transition of the pure gauge theory is recovered at $G_s=6.3$ GeV$^{-2}$,
only 15\% above the reference value \eqref{eq:coupling_physical}.

One may wonder why this mechanism is not effective for the
polynomial potential, since it also exhibits a first order phase
transition in the absence of fermions. In fact, we find that the
polynomial potential does show the same behavior, but at much larger values of the scalar
coupling. The RW endpoint, which is second order at $G_s=5.498$
GeV$^{-2}$, is first order transition starting at $G_s=12.4$
GeV$^{-2}$ and the deconfinement CEP reaches $\mu^2=0$ at $G_s=25$ GeV$^{-2}$, 4.5
times larger than the reference value. 

The origin of this quantitative
difference between the polynomial and logarithmic potentials is due to 
the much weaker first order phase transition (smaller discontinuity in $\Phi$)
exhibited by the polynomial potential in the heavy quark limit.
At  $G_s=25$ GeV$^{-2}$ we find a dynamical quark mass of about 2.5 GeV. 
Thus, for a quark mass less than 2.5 GeV the first order transition is smoothened 
to a crossover transition. By contrast, for the logarithmic
potential this happens at a much smaller dynamical quark mass of 0.4 GeV, owing to the 
much stronger underlying first order transition. 

We note that a first order
confinement-deconfinement transition emerges at real chemical potential 
also in the large $N_c$ limit of the PNJL model
\cite{mclerran09:_quark_matter_and_chiral_symmet_break}, as explored  in the context
of the recently proposed quarkyonic phase
\cite{mclerran07:_phases_of_dense_quark_at}. 
Indeed, the effect of strengthening $G_s$ is similar to that of increasing $N_c$
since $G_s$ and $N_{c}$ appear  in the factor $G_sN_c$  in the gap equation for
the dynamical mass (see Eq.(16) in Ref.~\cite{mclerran09:_quark_matter_and_chiral_symmet_break}).
While we suppress the quark
contribution to the thermodynamics by increasing the dynamical mass by means of
$G_s$, a large value of $N_c$ is accompanied by a $1/N_c$ suppression of the quark
contribution. Both procedures yield
a gluon dominated system and thus give a first order
confinement-deconfinement transition.

Note, however, that the two procedures differ in detail.
Increasing $G_{s}$ at fixed $N_{c}$ preserves the Polyakov loop potential but modifies the quark mass, 
while in the large $N_{c}$ limit at fixed $G_{s}N_{c}$ the quark mass remains unchanged but the 
Polyakov loop potential is modified. This means that when we increase $G_{s}$ at fixed $N_{c}$ we change
the characteristic scale of the chiral symmetry breaking. Although this does not correspond to the physical situation,
since QCD has unique scale, $\Lambda_{\text{QCD}}$, our result could be useful for exploring the interplay between the chiral
phase transition and deconfinement.

\begin{figure}[!t]
 \includegraphics[width=3.375in]{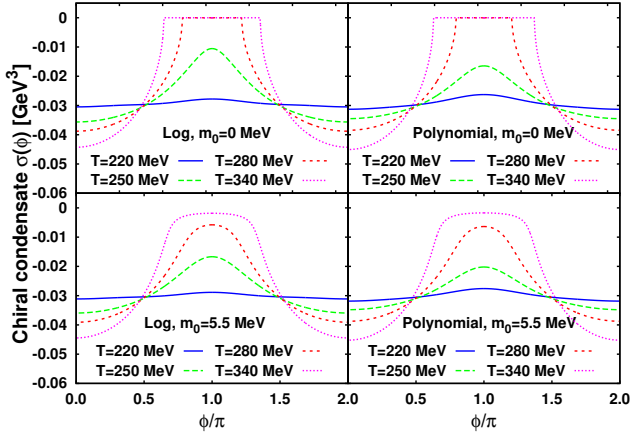}
 \caption{Chiral condensate $\sigma=\langle \bar{q}q \rangle$ as a function of
 the twisted angle $\varphi$.}
 \label{fig:sigma-varphi}
\end{figure}

\section{Dual parameter for the confinement-deconfinement transition}
\label{sec:dual}

\begin{figure}[ht]
 \includegraphics[width=3.375in]{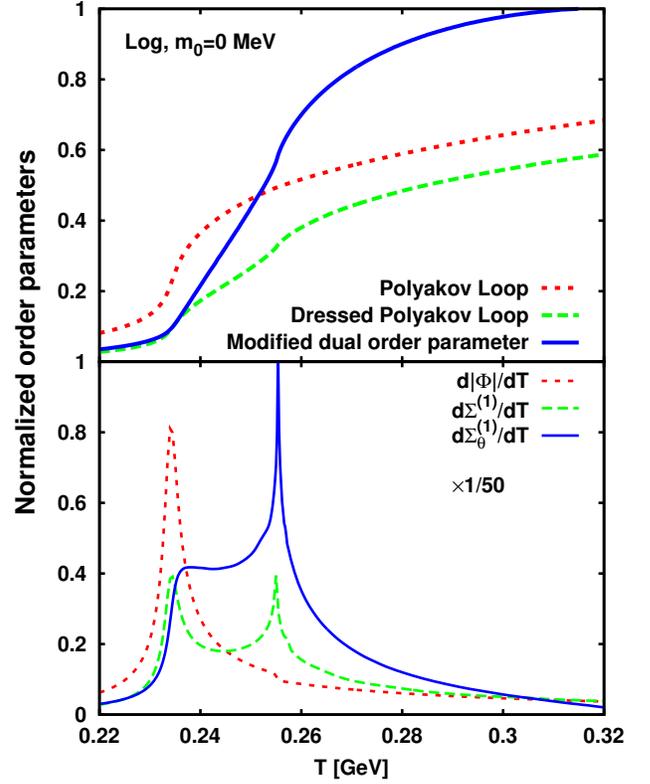}
 \caption{Characteristic parameters (upper) and their derivatives with respect to
 temperature (lower) for the confinement-deconfinement
 transition in the chiral limit for the logarithmic potential. }
 \label{fig:dual_n1_log}
\end{figure}

\begin{figure}[ht]
 \includegraphics[width=3.375in]{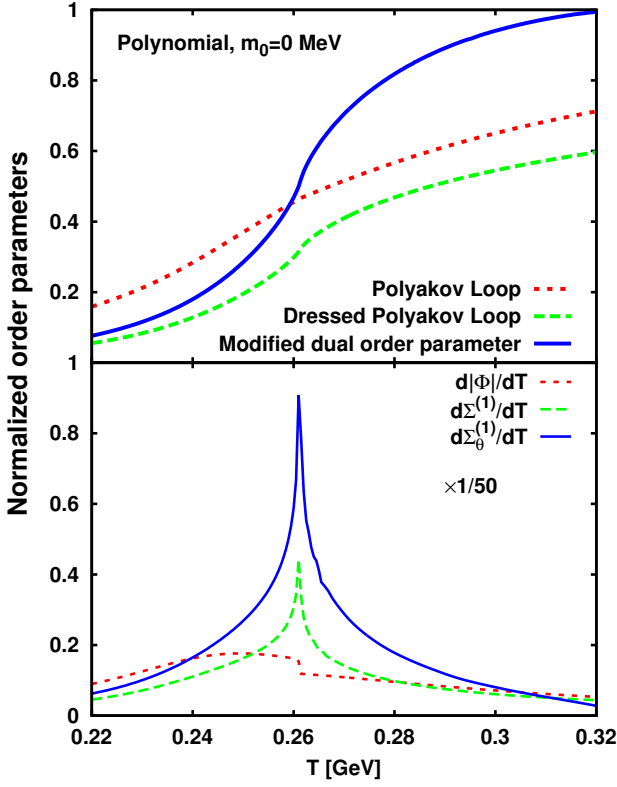}
 \caption{Same as Fig.~\ref{fig:dual_n1_log}, but for the polynomial potential.}
 \label{fig:dual_n1_poly}
\end{figure}

In this section, we consider dual parameters which capture
the characteristic feature of the confinement-deconfinement transition
discussed above.
Recently a dual parameter has been introduced by considering a generalized boundary
condition for fermions
\begin{equation}
 q(\boldsymbol{x},\beta) = e^{i\varphi}q(\boldsymbol{x},0).\label{eq:twistedboundary}
\end{equation}
Here $\varphi$ is so-called the twisted angle. The dual quark condensate 
$\Sigma^{(n)}$ is defined as the $n$-th
Fourier coefficient of the chiral condensate as a function of the
twisted angle \cite{bilgici08:_dual_polyak}
\begin{equation}
 \Sigma^{(n)}(T)=-\int_{0}^{2\pi}\frac{d\varphi}{2\pi} e^{-in\varphi}\sigma(T,\varphi).\label{eq:dual_twist}
\end{equation}
The chiral condensate $\sigma(T,\varphi)$ is defined in terms of $\varphi$
by \cite{bilgici08:_dual_polyak}
\begin{equation}
 \sigma(\varphi) = -\frac{1}{V}\langle
  \text{Tr}[(m_0+D_\varphi)^{-1}]\rangle_G .\label{eq:sigma_twist}
\end{equation}
In Ref.~\cite{bilgici08:_dual_polyak}, this quantity was introduced
based on the lattice regularization. The $\varphi$ dependence of 
$\text{Tr}[(m_0+D_\varphi)^{-1}]$  can be written down explicitly by using
the link variable. 
It reduces to the ordinary chiral order parameter in the limit
of $m\rightarrow 0$ and $V\rightarrow\infty$. 
An implementation in the continuum theory has been
done in the framework of Dyson-Schwinger equation
\cite{fischer09:_chiral_dyson_schwin}.
The most interesting quantity is that of $n=1$, which is called dressed
Polyakov loop. Because of a relation to the ordinary Polyakov loop,
it can be regarded as an order parameter of the
confinement-deconfinement transition.

\begin{figure}[!t]
 \includegraphics[width=3.375in]{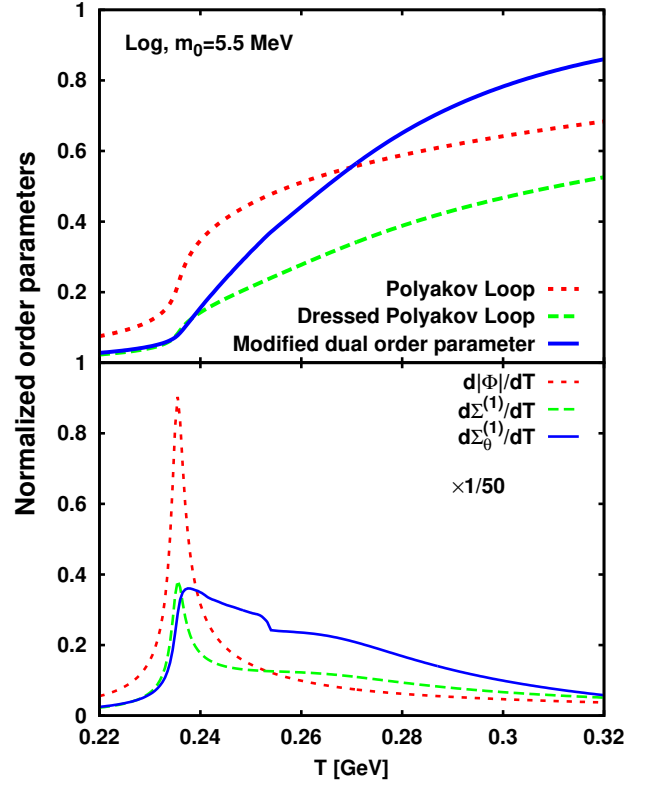}
 \caption{Same as Fig.~\ref{fig:dual_n1_log}, but for $m_0=5.5$ MeV.}
 \label{fig:dual_n1_m55log}
\end{figure}
\begin{figure}[!t]
 \includegraphics[width=3.375in]{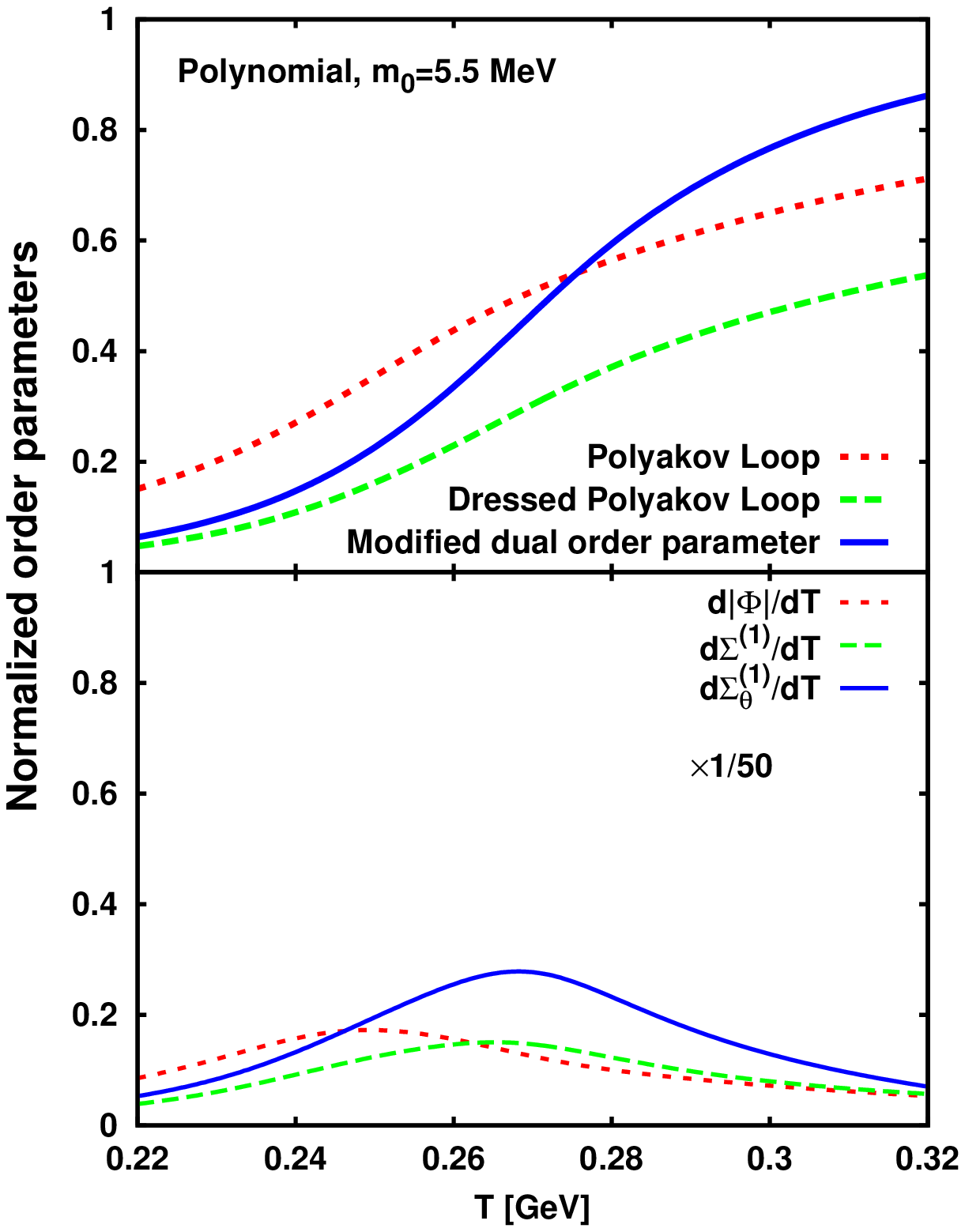}
 \caption{Same as Fig.~\ref{fig:dual_n1_poly}, but for $m_0=5.5$ MeV.}
 \label{fig:dual_n1_m55poly}
\end{figure}

From the fermionic boundary conditions
\eqref{eq:twistedboundary}, one immediately finds that this is
equivalent to introducing the imaginary chemical potential
[Eq.~\eqref{eq:boundary_immu}]. The only difference is that
$\varphi=\pm \pi$ corresponds to the usual anti-periodic boundary
condition in the twisted angle while $\theta=0$ does so in the imaginary
chemical potential. In this case, the relation between $\varphi$ and
$\theta$ is given by just a shift of $\pm\pi$,
\begin{equation}
 \varphi = \theta \pm \pi \quad (\text{mod} \quad 2\pi/N_c)\label{eq:twisted-imaginarymu}.
\end{equation}
Furthermore, one notes that the dual condensate \eqref{eq:dual_twist} is
quite similar to the canonical partition function \eqref{eq:canonical}. 

However, the LGT calculations demonstrate that
 $\sigma(\varphi)$ exhibits quite different behavior from that for
the imaginary chemical potential \cite{bilgici10:_fermion_qcd}. 
$\sigma(\varphi)$ shows a periodicity
of $2\pi$ in $\varphi$, not $2\pi/3$ which is required by the RW
periodicity.  The origin of this difference is the expectation value of
the operator $\langle\cdots \rangle_G$ in Eq.~\eqref{eq:sigma_twist}.
The subscript $G$ denotes the path integral over the gauge field with the
fermion determinant which follows the ordinary boundary condition. 
The change of the boundary condition \eqref{eq:twistedboundary} applies
only to the Dirac operator. In the case of the imaginary chemical
potential, different $\theta$ gives a different fermion determinant
while in the case of the twisted angle the background field does not
change with the boundary condition. Therefore $\varphi$ dependence of
the chiral condensate differs from $\theta$ dependence. 
In a PNJL model, the authors
of Ref.~\cite{kashiwa09:_dual_polyak_nambu_jona_lasin} use the value of
the Polyakov loop calculated at $\theta=0$ to
obtain the chiral condensate $\sigma(\varphi)$. 
This prescription corresponds to varying only the fermionic boundary
 condition without changing gluonic background. 
We will follow the same prescription below.
Since the periodicity $2\pi/3$ in the imaginary chemical potential is
preserved by the RW transition, which is an effect of the Polyakov loop
in the context of PNJL model, the relation
\eqref{eq:twisted-imaginarymu} holds for the normal NJL model
calculation which does not couple to $Z(3)$ field. In spite of the
absence of confinement in the NJL model, one sees that behavior of
the dual chiral condensate is quite similar to one obtained from lattice
QCD and PNJL model \cite{mukherjee:_chiral_polyak_nambu_jona_lasin}.

Figure \ref{fig:sigma-varphi} shows the chiral condensate as a function
of the twisted angle $\varphi$, obtained by the same method as used in
Ref.~\cite{kashiwa09:_dual_polyak_nambu_jona_lasin}. The
periodicity is  no longer $2\pi/3$. One also notes that there is broken
phase at region far from $\varphi=\pi$ even at high temperatures. This
is in contrast to the case of the imaginary chemical potential
shown in Fig.~\ref{fig:sigma-theta_m0}
and similar to what was expected from the gap equation of the NJL model,
Eq.~\eqref{eq:gapeq_NJL_boltzmann}. Indeed, at $\varphi=\pi/2$
and $3\pi/2$, which correspond to $\theta=\mp\pi/2$, the  thermal term
vanishes in Eq.~\eqref{eq:gapeq_NJL_boltzmann}, resulting in the almost
temperature indepedent chiral condensate.
On the other hand, there are no 
qualitative difference between the logarithmic potential and the
polynomial one.  The reason is   that the Polyakov
loop enters in the gap equation only as a constant determined at
$\varphi=\pi$ at each temperatures. 

Let us introduce a new dual parameter by using $\theta$ instead of
$\varphi$ such that it captures the characteristics in the $\theta$
space. We define
\begin{equation}
 \Sigma_\theta^{(n)}(T) \equiv \frac{3}{2\pi}\intop_{-\pi/3}^{\pi/3}d\theta e^{-in\theta}\sigma(T,\theta)\label{eq:dual_modified}.
\end{equation}
The integration region is changed to $[-\pi/3:\pi/3]$ with respect to
the periodicity $2\pi/3$. As discussed in
Sec.~\ref{sec:analyticinsight}, the physical meaning of the periodicity
is different in confinement and deconfinement phase.
In the confinement phase, periodicity $2\pi/3$ is coming from
$\cos3\theta$ which characterizes the confinement of the quarks. 
On the other hand, deconfinement phase is characterized by $\cos\theta$
with discontinuity at $\theta=\pi/3 +2\pi k/3$ caused by $Z(3)$
transition which preserves the periodicity $2\pi/3$.
Therefore, we expect, that $\Sigma_\theta^{(1)}$ and $\Sigma_\theta^{(3)}$
demonstrate characteristic behavior of the confinement-deconfinement transition.

\begin{figure}[htbp]
 \includegraphics[width=3.375in]{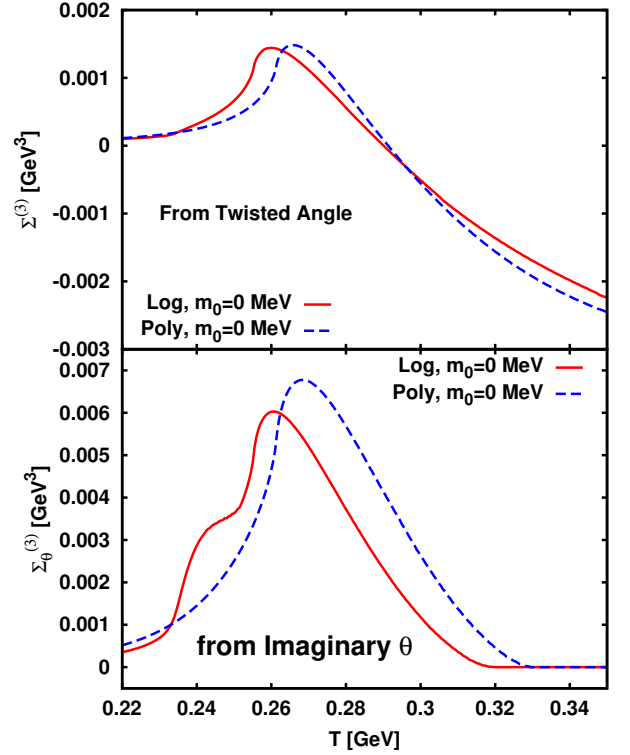}
 \caption{Dual order parameters of $n=3$. Upper panel shows the result
 from Eq.~\eqref{eq:dual_twist} while lower one displays that from
 Eq.~\eqref{eq:dual_modified}.}
 \label{fig:n3dual}
\end{figure}

In Figs.~\ref{fig:dual_n1_log} and \ref{fig:dual_n1_poly}, we compare the three kinds of the characteristic
parameters of the confinement-deconfinement transition and their
derivatives with respect to temperature as functions of temperature. 
We normalized the dressed Polyakov loop and the modified dual
parameter \eqref{eq:dual_modified} so that they tend to 0 at low
temperature and to unity at high temperature by
$\Sigma_\text{norm}(T)\equiv [\Sigma(T)-\Sigma(T_1)]/[\Sigma(T_2)-\Sigma(T_1)]$ 
where we used $T_1=50$ MeV and $T_2=1$ GeV.\footnote{Note that $\Sigma_\theta^{(1)}(T=0)$ does not
vanish since the integration is from $-\pi/3$ to $\pi/3$, not from
$-\pi$ to $\pi$.} 
The same normalization constants are applied to the derivatives.
Note that $\Sigma_\theta^{(n)}$ before normalization vanishes
at temperature higher than the chiral transition temperature at
$\theta=\pi/3$ in the chiral limit since $\sigma(T,\theta)$ does so, as seen in
Fig.~\ref{fig:sigma-theta_m0}. After the normalization, it smoothly
approaches to unity as one sees in
Figs.~\ref{fig:dual_n1_log}--\ref{fig:dual_n1_m55poly}. 

One observes qualitatively that both dual parameters
show similar behavior to the Polyakov loop.
The dressed Polyakov loop is almost parallel to the
Polyakov loop in the temperature region covered in the figure.
On the other hand, the modified dual
parameter $\Sigma_\theta^{(1)}$ reaches quickly its limiting value. 
The behavior around the transition temperatures reflect
the difference between two potentials. 
In the case of the logarithmic potential, one sees different structures
for each derivative of the order parameters.
While the derivative of the Polyakov loop $d|\Phi|/dT$ shows the only
peak corresponding to the pseudocritical temperature, the
dressed Polyakov loop $d\Sigma^{1}/dT$ exhibits the two peak structure. 
One agrees with that of the $d\Phi/dT$ and the other corresponds to the
chiral transition (see Fig.~\ref{fig:mu0}). The derivative of the modified dual 
parameter $d\Sigma_\theta^{(1)}/dT$ also shows a peak for the chiral
transition. However, the deconfinement appears only as a shoulder. The
polynomial potential shows a broader peak in $d\Phi/dT$
reflecting the weaker nature of transition. However, maxima in dual
order parameters associated with the crossover transition do not appear.

Inclusion of small quark mass, $m_0=5.5$ MeV slightly modifies the
behavior of the characteristic parameters, as expected from the
difference in $\sigma$. Figures \ref{fig:dual_n1_m55log} and
\ref{fig:dual_n1_m55poly} show the three parameters and their
derivatives for $m_0=5.5$ MeV. While at finite quark mass there is
little difference in the behavior of the parameters seen in $m_0=0$
case, distinct peak structures 
show up in their derivatives. For the logarithmic potential, the peak
associated with the chiral transition seen in the chiral limit does not
exist in both dual parameters. The remnant of the chiral transition
appears only in the modified dual parameter as a dip. On the other hand,
polynomial potential case exhibits much broader peak, which
corresponds to deconfinement in $d|\Phi|/dT$ and to chiral transition in
dual parameters.

Figure \ref{fig:n3dual} shows the dual parameter \eqref{eq:dual_twist}
and \eqref{eq:dual_modified} at $n=3$.
In all the cases, one sees that they start to increase, exhibit peak
structure and then decreasing. This is common for both $\Sigma^{(3)}$
and $\Sigma_\theta^{(3)}$.
Such behavior of $\Sigma_\theta^{(3)}$ can be
understood by analysing  
$\sigma(T,\theta)$ which is shown in Fig.~\ref{fig:sigma-theta_m0}.
At low $T$, it oscillates according to $\cos3\theta$ with
an amplitude given by the thermal factor. Then $\Sigma_\theta^{(3)}$, which is
proportional to the amplitude, increases with $T$. At $T > T_\chi$, 
$\sigma(\theta,T)$ vanishes at $\theta=[0:\theta_c]$ thus only the
region $[\theta_c:\pi/3]$ contributes to the integral.
Therefore, $\Sigma_\theta^{(3)}$ has a peak at $T=T_\chi$ and then 
decreases. This does not relate to the deconfinement phenomenon and is
common for $\Sigma^{(3)}$ and $\Sigma_\theta^{(3)}$.

If one focuses on the difference between the logarithmic an the
polynomial potential, however, one finds a remnant of the
deconfinement-confinement transition in the behavior of
$\Sigma_\theta^{(3)}$. As we have seen in Fig.~\ref{fig:firstorder},
$\sigma(T,\theta)$ has a discontinuity 
induced by the first-order confinement-deconfinement transition.
It is reflected to the non-monotonic
behavior of $\Sigma_\theta^{(3)}$ between $T=0.24$ and $T=0.26$ GeV in the case of
logarithmic potential. Indeed, the shoulder seen at $T=0.24$ GeV
corresponds to the temperature of CEP, at which $\sigma(\theta,T)$ has a
discontinuity, and the second inflection point reflects the RW endpoint
which is the triple point in this case.
This indicates that the newly introduced dual parameter
$\Sigma_\theta^{(3)}$ has sensitivity to the confinement-deconfinement
transition at imaginary chemical potential.

\section{Summary}
\label{sec:summary}

We have studied the confinement-deconfinement transition in the PNJL
model at imaginary chemical potential with the simplest interaction. 
We discussed the origin of the characteristic periodicity $2\pi/3$ of
the order parameters. 
It is characterized by $\cos3\theta$ in the confined phase while
it is due to $\cos\theta$ with the RW transition at
$\theta=\pi/3+2\pi k/3$ induced by the change of the phase of the
Polyakov loop. We also explored the results from different
Polyakov loop potentials. We found that the property of the
confinement-deconfinement transition depends on the choice of the potential
in spite of the fact that both potentials exhibit the first-order
phase transition in the absence of
quarks. Remarkable differences are seen in both the RW endpoint and the
behavior of the Polyakov loop at finite $\theta$. In the case of the
logarithmic potential which has more relevant domain, we find that the 
confinement-deconfinement transition becomes first order near
$\theta=\pi/3$ and there is a critical endpoint of the transition at
imaginary chemical potential. We also find that the location of the CEP 
moves with the four fermion coupling $G_s$ and it reaches real chemical
potential region by increasing $G_s$. This behavior
can be understood by the suppression of the quark contribution since
increasing $G_s$ implies larger $\langle \bar{q}q \rangle$ which
quantifies the dynamical quark mass. At large coupling, the
existence of the CEP is independent of the choice of the
potential. However, the polynomial potential requires larger $G_s$
because it exhibits much weaker first order transition.
Consequently, it seems that the order of the deconfinement
transition is determined by the size of the gap $\Delta\Phi$ in the
Polyakov loop potential and the quark condensate $\langle \bar{q}q \rangle$.

The first order phase transition influences the behavior of the chiral
condensate as a sudden jump at the critical imaginary chemical potential. 
We proposed modified dual parameters using the imaginary
chemical potential based on the analogy to the twisted angle in the dual order
parameters. Comparing the $n=1$ case with the Polyakov loop and the
dressed Polyakov loop, we found that each parameters has different
sensitivity to the phase transitions. 
We showed that $n=3$ has a characteristic behavior owing to the
first order confinement-deconfinement transition at intermediate
$\theta$. We expect relevance of our study in understanding the QCD phase diagram.

\begin{acknowledgments}
 K.M. would like to thank Y. Sakai, T. Sasaki, and M. Yahiro for
 fruitful discussion and warm hospitality during his visit to Kyushu University.
 K.M. and V.S. would like to acknowledge Frankfurt institute of advanced study
 (FIAS) for support. 
 K.M. is supported by Yukawa International Program for Quark Hadron
 Sciences at Kyoto University.
 B.F. and K.R. acknowledges partial support by EMMI.
 K.R. acknowledges partial support by the Polish Ministry of Science (MEN)
\end{acknowledgments}

%


\begin{thebibliography}{54}%
\makeatletter
\providecommand \@ifxundefined [1]{%
 \@ifx{#1\undefined}
}%
\providecommand \@ifnum [1]{%
 \ifnum #1\expandafter \@firstoftwo
 \else \expandafter \@secondoftwo
 \fi
}%
\providecommand \@ifx [1]{%
 \ifx #1\expandafter \@firstoftwo
 \else \expandafter \@secondoftwo
 \fi
}%
\providecommand \natexlab [1]{#1}%
\providecommand \enquote  [1]{``#1''}%
\providecommand \bibnamefont  [1]{#1}%
\providecommand \bibfnamefont [1]{#1}%
\providecommand \citenamefont [1]{#1}%
\providecommand \href@noop [0]{\@secondoftwo}%
\providecommand \href [0]{\begingroup \@sanitize@url \@href}%
\providecommand \@href[1]{\@@startlink{#1}\@@href}%
\providecommand \@@href[1]{\endgroup#1\@@endlink}%
\providecommand \@sanitize@url [0]{\catcode `\\12\catcode `\$12\catcode
  `\&12\catcode `\#12\catcode `\^12\catcode `\_12\catcode `\%12\relax}%
\providecommand \@@startlink[1]{}%
\providecommand \@@endlink[0]{}%
\providecommand \url  [0]{\begingroup\@sanitize@url \@url }%
\providecommand \@url [1]{\endgroup\@href {#1}{\urlprefix }}%
\providecommand \urlprefix  [0]{URL }%
\providecommand \Eprint [0]{\href }%
\providecommand \doibase [0]{http://dx.doi.org/}%
\providecommand \selectlanguage [0]{\@gobble}%
\providecommand \bibinfo  [0]{\@secondoftwo}%
\providecommand \bibfield  [0]{\@secondoftwo}%
\providecommand \translation [1]{[#1]}%
\providecommand \BibitemOpen [0]{}%
\providecommand \bibitemStop [0]{}%
\providecommand \bibitemNoStop [0]{.\EOS\space}%
\providecommand \EOS [0]{\spacefactor3000\relax}%
\providecommand \BibitemShut  [1]{\csname bibitem#1\endcsname}%
\let\auto@bib@innerbib\@empty
\bibitem [{\citenamefont {Bazavov}\ \emph {et~al.}(2009)\citenamefont {Bazavov}
  \emph {et~al.}}]{bazavov09:_equat_qcd}%
  \BibitemOpen
  \bibfield  {author} {\bibinfo {author} {\bibfnamefont {A.}~\bibnamefont
  {Bazavov}} \emph {et~al.},\ }\href@noop {} {\bibfield  {journal} {\bibinfo
  {journal} {Phys. Rev. D}\ }\textbf {\bibinfo {volume} {80}},\ \bibinfo
  {pages} {014504} (\bibinfo {year} {2009})}\BibitemShut {NoStop}%
\bibitem [{\citenamefont {Aoki}\ \emph {et~al.}(2009)\citenamefont {Aoki},
  \citenamefont {Bors\'{a}nyi}, \citenamefont {D\"{u}rr}, \citenamefont
  {Fodor}, \citenamefont {Katz}, \citenamefont {Krieg},\ and\ \citenamefont
  {Szabo}}]{aoki09:_qcd}%
  \BibitemOpen
  \bibfield  {author} {\bibinfo {author} {\bibfnamefont {Y.}~\bibnamefont
  {Aoki}}, \bibinfo {author} {\bibfnamefont {S.}~\bibnamefont {Bors\'{a}nyi}},
  \bibinfo {author} {\bibfnamefont {S.}~\bibnamefont {D\"{u}rr}}, \bibinfo
  {author} {\bibfnamefont {Z.}~\bibnamefont {Fodor}}, \bibinfo {author}
  {\bibfnamefont {S.~D.}\ \bibnamefont {Katz}}, \bibinfo {author}
  {\bibfnamefont {S.}~\bibnamefont {Krieg}}, \ and\ \bibinfo {author}
  {\bibfnamefont {K.}~\bibnamefont {Szabo}},\ }\href@noop {} {\bibfield
  {journal} {\bibinfo  {journal} {JHEP}\ }\textbf {\bibinfo {volume} {0906}},\
  \bibinfo {pages} {088} (\bibinfo {year} {2009})}\BibitemShut {NoStop}%
\bibitem{fodor}
	Y.~Aoki, G.~Endr\"{o}di, Z.~Fodor, S.~D.~Katz, and K.~K.~Szabo,
	Nature \textbf{443} (2006) 675.

\bibitem [{\citenamefont {Muroya}\ \emph {et~al.}(2003)\citenamefont {Muroya},
  \citenamefont {Nakamura}, \citenamefont {Nonaka},\ and\ \citenamefont
  {Takaishi}}]{muroya03:_lattic_qcd}%
  \BibitemOpen
  \bibfield  {author} {\bibinfo {author} {\bibfnamefont {S.}~\bibnamefont
  {Muroya}}, \bibinfo {author} {\bibfnamefont {A.}~\bibnamefont {Nakamura}},
  \bibinfo {author} {\bibfnamefont {C.}~\bibnamefont {Nonaka}}, \ and\ \bibinfo
  {author} {\bibfnamefont {T.}~\bibnamefont {Takaishi}},\ }\href@noop {}
  {\bibfield  {journal} {\bibinfo  {journal} {Prog. Theor. Phys.}\ }\textbf
  {\bibinfo {volume} {110}},\ \bibinfo {pages} {615} (\bibinfo {year}
  {2003})}\BibitemShut {NoStop}%
\bibitem [{\citenamefont {Roberge}\ and\ \citenamefont
  {Weiss}(1986)}]{roberge86:_gauge_qcd}%
  \BibitemOpen
  \bibfield  {author} {\bibinfo {author} {\bibfnamefont {A.}~\bibnamefont
  {Roberge}}\ and\ \bibinfo {author} {\bibfnamefont {N.}~\bibnamefont
  {Weiss}},\ }\href@noop {} {\bibfield  {journal} {\bibinfo  {journal} {Nucl.
  Phys.}\ }\textbf {\bibinfo {volume} {B275}},\ \bibinfo {pages} {734}
  (\bibinfo {year} {1986})}\BibitemShut {NoStop}%
\bibitem [{\citenamefont {Ejiri}(2008)}]{ejiri08:_canon_qcd}%
  \BibitemOpen
  \bibfield  {author} {\bibinfo {author} {\bibfnamefont {S.}~\bibnamefont
  {Ejiri}},\ }\href@noop {} {\bibfield  {journal} {\bibinfo  {journal} {Phys.
  Rev. D}\ }\textbf {\bibinfo {volume} {78}},\ \bibinfo {pages} {074507}
  (\bibinfo {year} {2008})}\BibitemShut {NoStop}%
\bibitem [{\citenamefont {Li}\ \emph {et~al.}(2010)\citenamefont {Li},
  \citenamefont {Alexandru}, \citenamefont {Liu},\ and\ \citenamefont
  {Meng}}]{li:_finit_qcd_n_n}%
  \BibitemOpen
  \bibfield  {author} {\bibinfo {author} {\bibfnamefont {A.}~\bibnamefont
  {Li}}, \bibinfo {author} {\bibfnamefont {A.}~\bibnamefont {Alexandru}},
  \bibinfo {author} {\bibfnamefont {K.~F.}\ \bibnamefont {Liu}}, \ and\
  \bibinfo {author} {\bibfnamefont {X.}~\bibnamefont {Meng}},\ }\href@noop {}
  {\bibfield  {journal} {\bibinfo  {journal} {Phys. Rev. D}\ }\textbf {\bibinfo
  {volume} {82}},\ \bibinfo {pages} {054502} (\bibinfo {year} {2010})},\
  \Eprint {http://arxiv.org/abs/arXiv:1005.4158} {arXiv:1005.4158} \BibitemShut
  {NoStop}%
\bibitem [{\citenamefont {de~Forcrand}\ and\ \citenamefont
  {Philipsen}(2002)}]{forcrand02:_qcd}%
  \BibitemOpen
  \bibfield  {author} {\bibinfo {author} {\bibfnamefont {P.}~\bibnamefont
  {de~Forcrand}}\ and\ \bibinfo {author} {\bibfnamefont {O.}~\bibnamefont
  {Philipsen}},\ }\href@noop {} {\bibfield  {journal} {\bibinfo  {journal}
  {Nucl. Phys.}\ }\textbf {\bibinfo {volume} {B642}},\ \bibinfo {pages} {290}
  (\bibinfo {year} {2002})}\BibitemShut {NoStop}%
\bibitem [{\citenamefont {de~Forcrand}\ and\ \citenamefont
  {Philipsen}(2003)}]{forcrand03:_qcd}%
  \BibitemOpen
  \bibfield  {author} {\bibinfo {author} {\bibfnamefont {P.}~\bibnamefont
  {de~Forcrand}}\ and\ \bibinfo {author} {\bibfnamefont {O.}~\bibnamefont
  {Philipsen}},\ }\href@noop {} {\bibfield  {journal} {\bibinfo  {journal}
  {Nucl. Phys.}\ }\textbf {\bibinfo {volume} {B673}},\ \bibinfo {pages} {170}
  (\bibinfo {year} {2003})}\BibitemShut {NoStop}%
\bibitem [{\citenamefont {D'Elia}\ and\ \citenamefont
  {Lombardo}(2003)}]{d'elia03:_finit_qcd}%
  \BibitemOpen
  \bibfield  {author} {\bibinfo {author} {\bibfnamefont {M.}~\bibnamefont
  {D'Elia}}\ and\ \bibinfo {author} {\bibfnamefont {M.~P.}\ \bibnamefont
  {Lombardo}},\ }\href@noop {} {\bibfield  {journal} {\bibinfo  {journal}
  {Phys. Rev. D}\ }\textbf {\bibinfo {volume} {67}},\ \bibinfo {pages} {014505}
  (\bibinfo {year} {2003}).}

 \bibitem{Papa}
	 P.~Giudice and A.~Papa,
	 Phys.\ Rev.\ D \textbf{69}, 094509 (2004);  P.~Cea, L.~Cosmai,
	 M.~D'Elia, C.~Manneschi, and A.~Papa,
	 \textit{ibid}. \textbf{80}, 034501 (2009).

\bibitem [{\citenamefont {D'Elia}\ and\ \citenamefont
  {Lombardo}(2004)}]{d'elia04:_qcd_b}%
  \BibitemOpen
  \bibfield  {author} {\bibinfo {author} {\bibfnamefont {M.}~\bibnamefont
  {D'Elia}}\ and\ \bibinfo {author} {\bibfnamefont {M.~P.}\ \bibnamefont
  {Lombardo}},\ }\href@noop {} {\bibfield  {journal} {\bibinfo  {journal}
  {Phys. Rev. D}\ }\textbf {\bibinfo {volume} {70}},\ \bibinfo {pages} {074509}
  (\bibinfo {year} {2004})}\BibitemShut {NoStop}%
\bibitem [{\citenamefont {Chen}\ and\ \citenamefont
  {Luo}(2005)}]{chen05:_phase_qcd_wilson}%
  \BibitemOpen
  \bibfield  {author} {\bibinfo {author} {\bibfnamefont {H.~S.}\ \bibnamefont
  {Chen}}\ and\ \bibinfo {author} {\bibfnamefont {X.~Q.}\ \bibnamefont {Luo}},\
  }\href@noop {} {\bibfield  {journal} {\bibinfo  {journal} {Phys. Rev. D}\
  }\textbf {\bibinfo {volume} {72}},\ \bibinfo {pages} {034504} (\bibinfo
  {year} {2005})}\BibitemShut {NoStop}%
\bibitem [{\citenamefont {D'Elia}\ \emph {et~al.}(2007)\citenamefont {D'Elia},
  \citenamefont {Renzo},\ and\ \citenamefont {Lombardo}}]{d'elia07:_stron_qcd}%
  \BibitemOpen
  \bibfield  {author} {\bibinfo {author} {\bibfnamefont {M.}~\bibnamefont
  {D'Elia}}, \bibinfo {author} {\bibfnamefont {F.~D.}\ \bibnamefont {Renzo}}, \
  and\ \bibinfo {author} {\bibfnamefont {M.~P.}\ \bibnamefont {Lombardo}},\
  }\href@noop {} {\bibfield  {journal} {\bibinfo  {journal} {Phys. Rev. D}\
  }\textbf {\bibinfo {volume} {76}},\ \bibinfo {pages} {114509} (\bibinfo
  {year} {2007})}\BibitemShut {NoStop}%
\bibitem [{\citenamefont {de~Forcrand}\ and\ \citenamefont
  {Philipsen}(2007)}]{forcrand07:_n_qcd}%
  \BibitemOpen
  \bibfield  {author} {\bibinfo {author} {\bibfnamefont {P.}~\bibnamefont
  {de~Forcrand}}\ and\ \bibinfo {author} {\bibfnamefont {O.}~\bibnamefont
  {Philipsen}},\ }\href@noop {} {\bibfield  {journal} {\bibinfo  {journal}
  {JHEP}\ }\textbf {\bibinfo {volume} {0701}},\ \bibinfo {pages} {077}
  (\bibinfo {year} {2007})}\BibitemShut {NoStop}%
\bibitem [{\citenamefont {Wu}\ \emph {et~al.}(2007)\citenamefont {Wu},
  \citenamefont {Luo},\ and\ \citenamefont {Chen}}]{wu07:_phase_qcd_wilson}%
  \BibitemOpen
  \bibfield  {author} {\bibinfo {author} {\bibfnamefont {L.~K.}\ \bibnamefont
  {Wu}}, \bibinfo {author} {\bibfnamefont {X.~Q.}\ \bibnamefont {Luo}}, \ and\
  \bibinfo {author} {\bibfnamefont {H.~S.}\ \bibnamefont {Chen}},\ }\href@noop
  {} {\bibfield  {journal} {\bibinfo  {journal} {Phys. Rev. D}\ }\textbf
  {\bibinfo {volume} {76}},\ \bibinfo {pages} {034505} (\bibinfo {year}
  {2007})}\BibitemShut {NoStop}%
\bibitem [{\citenamefont {de~Forcrand}\ and\ \citenamefont
  {Philipsen}(2010)}]{forcrand:_const_qcd}%
  \BibitemOpen
  \bibfield  {author} {\bibinfo {author} {\bibfnamefont {P.}~\bibnamefont
  {de~Forcrand}}\ and\ \bibinfo {author} {\bibfnamefont {O.}~\bibnamefont
  {Philipsen}},\ }\href@noop {} {\bibfield  {journal} {\bibinfo  {journal}
  {Phys. Rev. Lett.}\ }\textbf {\bibinfo {volume} {105}},\ \bibinfo {pages}
  {152001} (\bibinfo {year} {2010})},\ \Eprint
  {http://arxiv.org/abs/arXiv:1004.3144} {arXiv:1004.3144} \BibitemShut
  {NoStop}%
\bibitem [{\citenamefont {Nagata}\ and\ \citenamefont
  {Nakamura}(2011)}]{nagata11:_imagin_chemic_poten_approac_for}%
  \BibitemOpen
  \bibfield  {author} {\bibinfo {author} {\bibfnamefont {K.}~\bibnamefont
  {Nagata}}\ and\ \bibinfo {author} {\bibfnamefont {A.}~\bibnamefont
  {Nakamura}},\ }\href@noop {} {\bibfield  {journal} {\bibinfo  {journal}
  {Phys. Rev. D}\ }\textbf {\bibinfo {volume} {83}},\ \bibinfo {pages} {114507}
  (\bibinfo {year} {2011})}\BibitemShut {NoStop}%
\bibitem [{\citenamefont {Hart}\ \emph {et~al.}(2001)\citenamefont {Hart},
  \citenamefont {Laine},\ and\ \citenamefont {Philipsen}}]{hart01:_testing}%
  \BibitemOpen
  \bibfield  {author} {\bibinfo {author} {\bibfnamefont {A.}~\bibnamefont
  {Hart}}, \bibinfo {author} {\bibfnamefont {M.}~\bibnamefont {Laine}}, \ and\
  \bibinfo {author} {\bibfnamefont {O.}~\bibnamefont {Philipsen}},\ }\href@noop
  {} {\bibfield  {journal} {\bibinfo  {journal} {Phys. Lett. B}\ }\textbf
  {\bibinfo {volume} {505}},\ \bibinfo {pages} {141} (\bibinfo {year}
  {2001})}\BibitemShut {NoStop}%
\bibitem [{\citenamefont {Bluhm}\ and\ \citenamefont
  {K\"{a}mpfer}(2008)}]{bluhm08:_quasiparticle}%
  \BibitemOpen
  \bibfield  {author} {\bibinfo {author} {\bibfnamefont {M.}~\bibnamefont
  {Bluhm}}\ and\ \bibinfo {author} {\bibfnamefont {B.}~\bibnamefont
  {K\"{a}mpfer}},\ }\href@noop {} {\bibfield  {journal} {\bibinfo  {journal}
  {Phys. Rev. D}\ }\textbf {\bibinfo {volume} {77}},\ \bibinfo {pages} {034004}
  (\bibinfo {year} {2008})}\BibitemShut {NoStop}%
\bibitem [{\citenamefont {Weiss}(1987)}]{weiss87:_how}%
  \BibitemOpen
  \bibfield  {author} {\bibinfo {author} {\bibfnamefont {N.}~\bibnamefont
  {Weiss}},\ }\href@noop {} {\bibfield  {journal} {\bibinfo  {journal} {Phys.
  Rev. D}\ }\textbf {\bibinfo {volume} {35}},\ \bibinfo {pages} {2495}
  (\bibinfo {year} {1987})}\BibitemShut {NoStop}%
\bibitem [{\citenamefont {Kouno}\ \emph {et~al.}(2009)\citenamefont {Kouno},
  \citenamefont {Sakai}, \citenamefont {Kashiwa},\ and\ \citenamefont
  {Yahiro}}]{kouno09:_rober_weiss}%
  \BibitemOpen
  \bibfield  {author} {\bibinfo {author} {\bibfnamefont {H.}~\bibnamefont
  {Kouno}}, \bibinfo {author} {\bibfnamefont {Y.}~\bibnamefont {Sakai}},
  \bibinfo {author} {\bibfnamefont {K.}~\bibnamefont {Kashiwa}}, \ and\
  \bibinfo {author} {\bibfnamefont {M.}~\bibnamefont {Yahiro}},\ }\href@noop {}
  {\bibfield  {journal} {\bibinfo  {journal} {J. Phys. G: Nucl. Part. Phys.}\
  }\textbf {\bibinfo {volume} {36}},\ \bibinfo {pages} {115010} (\bibinfo
  {year} {2009})}\BibitemShut {NoStop}%
\bibitem [{\citenamefont {D'Elia}\ and\ \citenamefont
  {Sanfilippo}(2009)}]{d'elia09:_order_rober_weiss_qcd}%
  \BibitemOpen
  \bibfield  {author} {\bibinfo {author} {\bibfnamefont {M.}~\bibnamefont
  {D'Elia}}\ and\ \bibinfo {author} {\bibfnamefont {F.}~\bibnamefont
  {Sanfilippo}},\ }\href@noop {} {\bibfield  {journal} {\bibinfo  {journal}
  {Phys. Rev. D}\ }\textbf {\bibinfo {volume} {80}},\ \bibinfo {pages}
  {111501(R)} (\bibinfo {year} {2009})}\BibitemShut {NoStop}%
\bibitem [{\citenamefont {Sakai}\ \emph
  {et~al.}(2010{\natexlab{a}})\citenamefont {Sakai}, \citenamefont {Sasaki},
  \citenamefont {Kouno},\ and\ \citenamefont {Yahiro}}]{sakai:_10063408}%
  \BibitemOpen
  \bibfield  {author} {\bibinfo {author} {\bibfnamefont {Y.}~\bibnamefont
  {Sakai}}, \bibinfo {author} {\bibfnamefont {T.}~\bibnamefont {Sasaki}},
  \bibinfo {author} {\bibfnamefont {H.}~\bibnamefont {Kouno}}, \ and\ \bibinfo
  {author} {\bibfnamefont {M.}~\bibnamefont {Yahiro}},\ }\href@noop {}
  {\bibfield  {journal} {\bibinfo  {journal} {Phys. Rev. D}\ }\textbf {\bibinfo
  {volume} {82}},\ \bibinfo {pages} {076003} (\bibinfo {year}
  {2010}{\natexlab{a}})},\ \Eprint {http://arxiv.org/abs/arXiv:1006.3408}
  {arXiv:1006.3408} \BibitemShut {NoStop}%
\bibitem [{\citenamefont {Bonati}\ \emph {et~al.}(2011)\citenamefont {Bonati},
  \citenamefont {Cossu}, \citenamefont {D'Elia},\ and\ \citenamefont
  {Sanfilippo}}]{bonati:_rober_weiss_endpoin_in_n_f_qcd}%
  \BibitemOpen
  \bibfield  {author} {\bibinfo {author} {\bibfnamefont {C.}~\bibnamefont
  {Bonati}}, \bibinfo {author} {\bibfnamefont {G.}~\bibnamefont {Cossu}},
  \bibinfo {author} {\bibfnamefont {M.}~\bibnamefont {D'Elia}}, \ and\ \bibinfo
  {author} {\bibfnamefont {F.}~\bibnamefont {Sanfilippo}},\ }\href@noop {}
  {\bibfield  {journal} {\bibinfo  {journal} {Phys. Rev. D}\ }\textbf {\bibinfo
  {volume} {83}},\ \bibinfo {pages} {054505} (\bibinfo {year} {2011})},\
  \Eprint {http://arxiv.org/abs/arXiv:1011.4515} {arXiv:1011.4515.}

 \bibitem{Aarts}
	 G.~Aarts, S.~P.~Kumar, and J.~Rafferty, JHEP \textbf{1007}, 056
	 (2010).

\bibitem [{\citenamefont {Fukushima}(2004)}]{fukushima04:_chiral_polyak}%
  \BibitemOpen
  \bibfield  {author} {\bibinfo {author} {\bibfnamefont {K.}~\bibnamefont
  {Fukushima}},\ }\href@noop {} {\bibfield  {journal} {\bibinfo  {journal}
  {Phys. Lett. B}\ }\textbf {\bibinfo {volume} {591}},\ \bibinfo {pages} {277}
  (\bibinfo {year} {2004})}\BibitemShut {NoStop}%
\bibitem [{\citenamefont {Ratti}\ \emph {et~al.}(2006)\citenamefont {Ratti},
  \citenamefont {Thaler},\ and\ \citenamefont {Weise}}]{ratti06:_phases_qcd}%
  \BibitemOpen
  \bibfield  {author} {\bibinfo {author} {\bibfnamefont {C.}~\bibnamefont
  {Ratti}}, \bibinfo {author} {\bibfnamefont {M.~A.}\ \bibnamefont {Thaler}}, \
  and\ \bibinfo {author} {\bibfnamefont {W.}~\bibnamefont {Weise}},\
  }\href@noop {} {\bibfield  {journal} {\bibinfo  {journal} {Phys. Rev. D}\
  }\textbf {\bibinfo {volume} {73}},\ \bibinfo {pages} {014019} (\bibinfo
  {year} {2006})}\BibitemShut {NoStop}%
\bibitem [{\citenamefont {Nambu}\ and\ \citenamefont
  {Jona-Lasinio}(1961{\natexlab{a}})}]{nambu61:_NJLI}%
  \BibitemOpen
  \bibfield  {author} {\bibinfo {author} {\bibfnamefont {Y.}~\bibnamefont
  {Nambu}}\ and\ \bibinfo {author} {\bibfnamefont {G.}~\bibnamefont
  {Jona-Lasinio}},\ }\href@noop {} {\bibfield  {journal} {\bibinfo  {journal}
  {Phys. Rev.}\ }\textbf {\bibinfo {volume} {122}},\ \bibinfo {pages} {345}
  (\bibinfo {year} {1961}{\natexlab{a}})}\BibitemShut {NoStop}%
\bibitem [{\citenamefont {Nambu}\ and\ \citenamefont
  {Jona-Lasinio}(1961{\natexlab{b}})}]{nambu61:_NJLII}%
  \BibitemOpen
  \bibfield  {author} {\bibinfo {author} {\bibfnamefont {Y.}~\bibnamefont
  {Nambu}}\ and\ \bibinfo {author} {\bibfnamefont {G.}~\bibnamefont
  {Jona-Lasinio}},\ }\href@noop {} {\bibfield  {journal} {\bibinfo  {journal}
  {Phys. Rev.}\ }\textbf {\bibinfo {volume} {124}},\ \bibinfo {pages} {246}
  (\bibinfo {year} {1961}{\natexlab{b}})}\BibitemShut {NoStop}%
\bibitem [{\citenamefont {Hatsuda}\ and\ \citenamefont
  {Kunihiro}(1994)}]{hatsuda94:_qcd_lagran}%
  \BibitemOpen
  \bibfield  {author} {\bibinfo {author} {\bibfnamefont {T.}~\bibnamefont
  {Hatsuda}}\ and\ \bibinfo {author} {\bibfnamefont {T.}~\bibnamefont
  {Kunihiro}},\ }\href@noop {} {\bibfield  {journal} {\bibinfo  {journal}
  {Phys. Rept.}\ }\textbf {\bibinfo {volume} {247}},\ \bibinfo {pages} {221}
  (\bibinfo {year} {1994})}\BibitemShut {NoStop}%
\bibitem [{\citenamefont {Elze}\ \emph {et~al.}(1987)\citenamefont {Elze},
  \citenamefont {Miller},\ and\ \citenamefont {Redlich}}]{elze87:_gauge}%
  \BibitemOpen
  \bibfield  {author} {\bibinfo {author} {\bibfnamefont {H.~T.}\ \bibnamefont
  {Elze}}, \bibinfo {author} {\bibfnamefont {D.~E.}\ \bibnamefont {Miller}}, \
  and\ \bibinfo {author} {\bibfnamefont {K.}~\bibnamefont {Redlich}},\
  }\href@noop {} {\bibfield  {journal} {\bibinfo  {journal} {Phys. Rev. D}\
  }\textbf {\bibinfo {volume} {35}},\ \bibinfo {pages} {748} (\bibinfo {year}
  {1987})}\BibitemShut {NoStop}%
\bibitem [{\citenamefont {Miller}\ and\ \citenamefont
  {Redlich}(1988)}]{miller88:_therm_abelian}%
  \BibitemOpen
  \bibfield  {author} {\bibinfo {author} {\bibfnamefont {D.~E.}\ \bibnamefont
  {Miller}}\ and\ \bibinfo {author} {\bibfnamefont {K.}~\bibnamefont
  {Redlich}},\ }\href@noop {} {\bibfield  {journal} {\bibinfo  {journal} {Phys.
  Rev. D}\ }\textbf {\bibinfo {volume} {37}},\ \bibinfo {pages} {3716}
  (\bibinfo {year} {1988})}\BibitemShut {NoStop}%
\bibitem [{\citenamefont {Sakai}\ \emph
  {et~al.}(2008{\natexlab{a}})\citenamefont {Sakai}, \citenamefont {Kashiwa},
  \citenamefont {Kouno},\ and\ \citenamefont
  {Yahiro}}]{sakai08:_polyak_nambu_jona_lasin}%
  \BibitemOpen
  \bibfield  {author} {\bibinfo {author} {\bibfnamefont {Y.}~\bibnamefont
  {Sakai}}, \bibinfo {author} {\bibfnamefont {K.}~\bibnamefont {Kashiwa}},
  \bibinfo {author} {\bibfnamefont {H.}~\bibnamefont {Kouno}}, \ and\ \bibinfo
  {author} {\bibfnamefont {M.}~\bibnamefont {Yahiro}},\ }\href@noop {}
  {\bibfield  {journal} {\bibinfo  {journal} {Phys. Rev. D}\ }\textbf {\bibinfo
  {volume} {77}},\ \bibinfo {pages} {051901(R)} (\bibinfo {year}
  {2008}{\natexlab{a}})}\BibitemShut {NoStop}%
\bibitem [{\citenamefont {Sakai}\ \emph
  {et~al.}(2008{\natexlab{b}})\citenamefont {Sakai}, \citenamefont {Kashiwa},
  \citenamefont {Kouno},\ and\ \citenamefont {Yahiro}}]{sakai08:_phase_z}%
  \BibitemOpen
  \bibfield  {author} {\bibinfo {author} {\bibfnamefont {Y.}~\bibnamefont
  {Sakai}}, \bibinfo {author} {\bibfnamefont {K.}~\bibnamefont {Kashiwa}},
  \bibinfo {author} {\bibfnamefont {H.}~\bibnamefont {Kouno}}, \ and\ \bibinfo
  {author} {\bibfnamefont {M.}~\bibnamefont {Yahiro}},\ }\href@noop {}
  {\bibfield  {journal} {\bibinfo  {journal} {Phys. Rev. D}\ }\textbf {\bibinfo
  {volume} {78}},\ \bibinfo {pages} {036001} (\bibinfo {year}
  {2008}{\natexlab{b}})}\BibitemShut {NoStop}%
\bibitem [{\citenamefont {Sakai}\ \emph
  {et~al.}(2008{\natexlab{c}})\citenamefont {Sakai}, \citenamefont {Kashiwa},
  \citenamefont {Kouno}, \citenamefont {Matsuzaki},\ and\ \citenamefont
  {Yahiro}}]{sakai08:_vector_qcd}%
  \BibitemOpen
  \bibfield  {author} {\bibinfo {author} {\bibfnamefont {Y.}~\bibnamefont
  {Sakai}}, \bibinfo {author} {\bibfnamefont {K.}~\bibnamefont {Kashiwa}},
  \bibinfo {author} {\bibfnamefont {H.}~\bibnamefont {Kouno}}, \bibinfo
  {author} {\bibfnamefont {M.}~\bibnamefont {Matsuzaki}}, \ and\ \bibinfo
  {author} {\bibfnamefont {M.}~\bibnamefont {Yahiro}},\ }\href@noop {}
  {\bibfield  {journal} {\bibinfo  {journal} {Phys. Rev. D}\ }\textbf {\bibinfo
  {volume} {78}},\ \bibinfo {pages} {076007} (\bibinfo {year}
  {2008}{\natexlab{c}})}\BibitemShut {NoStop}%
\bibitem [{\citenamefont {Sakai}\ \emph {et~al.}(2009)\citenamefont {Sakai},
  \citenamefont {Kashiwa}, \citenamefont {Kouno}, \citenamefont {Matsuzaki},\
  and\ \citenamefont {Yahiro}}]{sakai09:_deter_qcd}%
  \BibitemOpen
  \bibfield  {author} {\bibinfo {author} {\bibfnamefont {Y.}~\bibnamefont
  {Sakai}}, \bibinfo {author} {\bibfnamefont {K.}~\bibnamefont {Kashiwa}},
  \bibinfo {author} {\bibfnamefont {H.}~\bibnamefont {Kouno}}, \bibinfo
  {author} {\bibfnamefont {M.}~\bibnamefont {Matsuzaki}}, \ and\ \bibinfo
  {author} {\bibfnamefont {M.}~\bibnamefont {Yahiro}},\ }\href@noop {}
  {\bibfield  {journal} {\bibinfo  {journal} {Phys. Rev. D}\ }\textbf {\bibinfo
  {volume} {79}},\ \bibinfo {pages} {096001} (\bibinfo {year}
  {2009})}\BibitemShut {NoStop}%
\bibitem [{\citenamefont {Sakai}\ \emph
  {et~al.}(2010{\natexlab{b}})\citenamefont {Sakai}, \citenamefont {Kouno},\
  and\ \citenamefont {Yahiro}}]{sakai:_09083088}%
  \BibitemOpen
  \bibfield  {author} {\bibinfo {author} {\bibfnamefont {Y.}~\bibnamefont
  {Sakai}}, \bibinfo {author} {\bibfnamefont {H.}~\bibnamefont {Kouno}}, \ and\
  \bibinfo {author} {\bibfnamefont {M.}~\bibnamefont {Yahiro}},\ }\href@noop {}
  {\bibfield  {journal} {\bibinfo  {journal} {J. Phys. G: Nucl. Part. Phys.}\
  }\textbf {\bibinfo {volume} {37}},\ \bibinfo {pages} {105007} (\bibinfo
  {year} {2010}{\natexlab{b}})},\ \Eprint
  {http://arxiv.org/abs/arXiv:0908.3088} {arXiv:0908.3088} \BibitemShut
  {NoStop}%
\bibitem [{\citenamefont {Kashiwa}\ \emph {et~al.}(2011)\citenamefont
  {Kashiwa}, \citenamefont {Hell},\ and\ \citenamefont
  {Weise}}]{kashiwa11:_nonloc_pnjl_model_and_imagin_chemic_poten}%
  \BibitemOpen
  \bibfield  {author} {\bibinfo {author} {\bibfnamefont {K.}~\bibnamefont
  {Kashiwa}}, \bibinfo {author} {\bibfnamefont {T.}~\bibnamefont {Hell}}, \
  and\ \bibinfo {author} {\bibfnamefont {W.}~\bibnamefont {Weise}},\
  }\href@noop {} \Eprint
  {http://arxiv.org/abs/arXiv:1106.5025} {arXiv:1106.5025} \BibitemShut
  {NoStop}%
\bibitem [{\citenamefont {Bilgici}\ \emph {et~al.}(2008)\citenamefont
  {Bilgici}, \citenamefont {Bruckmann}, \citenamefont {Gattringer},\ and\
  \citenamefont {Hagen}}]{bilgici08:_dual_polyak}%
  \BibitemOpen
  \bibfield  {author} {\bibinfo {author} {\bibfnamefont {E.}~\bibnamefont
  {Bilgici}}, \bibinfo {author} {\bibfnamefont {F.}~\bibnamefont {Bruckmann}},
  \bibinfo {author} {\bibfnamefont {C.}~\bibnamefont {Gattringer}}, \ and\
  \bibinfo {author} {\bibfnamefont {C.}~\bibnamefont {Hagen}},\ }\href@noop {}
  {\bibfield  {journal} {\bibinfo  {journal} {Phys. Rev. D}\ }\textbf {\bibinfo
  {volume} {77}},\ \bibinfo {pages} {094007} (\bibinfo {year}
  {2008})}\BibitemShut {NoStop}%
\bibitem [{\citenamefont {Sasaki}\ \emph {et~al.}(2007)\citenamefont {Sasaki},
  \citenamefont {Friman},\ and\ \citenamefont
  {Redlich}}]{sasaki07:_suscep_polyakov}%
  \BibitemOpen
  \bibfield  {author} {\bibinfo {author} {\bibfnamefont {C.}~\bibnamefont
  {Sasaki}}, \bibinfo {author} {\bibfnamefont {B.}~\bibnamefont {Friman}}, \
  and\ \bibinfo {author} {\bibfnamefont {K.}~\bibnamefont {Redlich}},\
  }\href@noop {} {\bibfield  {journal} {\bibinfo  {journal} {Phys. Rev. D}\
  }\textbf {\bibinfo {volume} {75}},\ \bibinfo {pages} {074013} (\bibinfo
  {year} {2007})}\BibitemShut {NoStop}%
\bibitem [{\citenamefont {Pisarski}(2000)}]{pisarski00:_quark_z_wilson}%
  \BibitemOpen
  \bibfield  {author} {\bibinfo {author} {\bibfnamefont {R.~D.}\ \bibnamefont
  {Pisarski}},\ }\href@noop {} {\bibfield  {journal} {\bibinfo  {journal}
  {Phys. Rev. D}\ }\textbf {\bibinfo {volume} {62}},\ \bibinfo {pages}
  {111501(R)} (\bibinfo {year} {2000})}\BibitemShut {NoStop}%
\bibitem [{\citenamefont {Boyd}\ \emph {et~al.}(1996)\citenamefont {Boyd},
  \citenamefont {Engles}, \citenamefont {Karsch}, \citenamefont {Laermann},
  \citenamefont {Legeland}, \citenamefont {L\"{u}tgemeier},\ and\ \citenamefont
  {Petersson}}]{Boyd_NPB469}%
  \BibitemOpen
  \bibfield  {author} {\bibinfo {author} {\bibfnamefont {G.}~\bibnamefont
  {Boyd}}, \bibinfo {author} {\bibfnamefont {J.}~\bibnamefont {Engles}},
  \bibinfo {author} {\bibfnamefont {F.}~\bibnamefont {Karsch}}, \bibinfo
  {author} {\bibfnamefont {E.}~\bibnamefont {Laermann}}, \bibinfo {author}
  {\bibfnamefont {C.}~\bibnamefont {Legeland}}, \bibinfo {author}
  {\bibfnamefont {M.}~\bibnamefont {L\"{u}tgemeier}}, \ and\ \bibinfo {author}
  {\bibfnamefont {B.}~\bibnamefont {Petersson}},\ }\href@noop {} {\bibfield
  {journal} {\bibinfo  {journal} {Nucl. Phys.}\ }\textbf {\bibinfo {volume}
  {B469}},\ \bibinfo {pages} {419} (\bibinfo {year} {1996})}\BibitemShut
  {NoStop}%
\bibitem [{\citenamefont {Kaczmarek}\ \emph {et~al.}(2002)\citenamefont
  {Kaczmarek}, \citenamefont {Karsch}, \citenamefont {Petreczky},\ and\
  \citenamefont {Zantow}}]{kaczmarek02:_heavy_quark_antiq_free_energ}%
  \BibitemOpen
  \bibfield  {author} {\bibinfo {author} {\bibfnamefont {O.}~\bibnamefont
  {Kaczmarek}}, \bibinfo {author} {\bibfnamefont {F.}~\bibnamefont {Karsch}},
  \bibinfo {author} {\bibfnamefont {P.}~\bibnamefont {Petreczky}}, \ and\
  \bibinfo {author} {\bibfnamefont {F.}~\bibnamefont {Zantow}},\ }\href@noop {}
  {\bibfield  {journal} {\bibinfo  {journal} {Phys. Lett. B}\ }\textbf
  {\bibinfo {volume} {543}},\ \bibinfo {pages} {41} (\bibinfo {year}
  {2002})}\BibitemShut {NoStop}%
\bibitem [{\citenamefont {Roessner}\ \emph {et~al.}(2007)\citenamefont
  {Roessner}, \citenamefont {Ratti},\ and\ \citenamefont
  {Weise}}]{roessner07:_polyak}%
  \BibitemOpen
  \bibfield  {author} {\bibinfo {author} {\bibfnamefont {S.}~\bibnamefont
  {Roessner}}, \bibinfo {author} {\bibfnamefont {C.}~\bibnamefont {Ratti}}, \
  and\ \bibinfo {author} {\bibfnamefont {W.}~\bibnamefont {Weise}},\
  }\href@noop {} {\bibfield  {journal} {\bibinfo  {journal} {Phys. Rev. D}\
  }\textbf {\bibinfo {volume} {75}},\ \bibinfo {pages} {034007} (\bibinfo
  {year} {2007})}.

 \bibitem{Fukushima2008} K.\ Fukushima, Phys.\ Rev.\ D \textbf{77},
	 114028 (2008).

\bibitem [{\citenamefont {Sasaki}\ and\ \citenamefont
  {Mishustin}(2010)}]{sasaki10:_therm_of_dense_hadron_matter}%
  \BibitemOpen
  \bibfield  {author} {\bibinfo {author} {\bibfnamefont {C.}~\bibnamefont
  {Sasaki}}\ and\ \bibinfo {author} {\bibfnamefont {I.}~\bibnamefont
  {Mishustin}},\ }\href@noop {} {\bibfield  {journal} {\bibinfo  {journal}
  {Phys. Rev. C}\ }\textbf {\bibinfo {volume} {82}},\ \bibinfo {pages} {035204}
  (\bibinfo {year} {2010})}\BibitemShut {NoStop}%
\bibitem [{\citenamefont {Wozar}\ \emph {et~al.}(2006)\citenamefont {Wozar},
  \citenamefont {Kaestner}, \citenamefont {Wipf}, \citenamefont {Heinzl},\ and\
  \citenamefont {Pozsgay}}]{wozar06:_phase_z_polyak}%
  \BibitemOpen
  \bibfield  {author} {\bibinfo {author} {\bibfnamefont {C.}~\bibnamefont
  {Wozar}}, \bibinfo {author} {\bibfnamefont {T.}~\bibnamefont {Kaestner}},
  \bibinfo {author} {\bibfnamefont {A.}~\bibnamefont {Wipf}}, \bibinfo {author}
  {\bibfnamefont {T.}~\bibnamefont {Heinzl}}, \ and\ \bibinfo {author}
  {\bibfnamefont {B.}~\bibnamefont {Pozsgay}},\ }\href@noop {} {\bibfield
  {journal} {\bibinfo  {journal} {Phys. Rev. D}\ }\textbf {\bibinfo {volume}
  {74}},\ \bibinfo {pages} {114501} (\bibinfo {year} {2006})}\BibitemShut
  {NoStop}%
\bibitem [{\citenamefont {Kondo}(2010)}]{kondo:_towar_qcd}%
  \BibitemOpen
  \bibfield  {author} {\bibinfo {author} {\bibfnamefont {K.~I.}\ \bibnamefont
  {Kondo}},\ }\href@noop {} {\bibfield  {journal} {\bibinfo  {journal} {Phys.
  Rev. D}\ }\textbf {\bibinfo {volume} {82}},\ \bibinfo {pages} {065024}
  (\bibinfo {year} {2010})},\ \Eprint {http://arxiv.org/abs/arXiv:1005.0314}
  {arXiv:1005.0314} \BibitemShut {NoStop}%
\bibitem [{\citenamefont {Braun}\ \emph {et~al.}(2011)\citenamefont {Braun},
  \citenamefont {Haas}, \citenamefont {Marhauser},\ and\ \citenamefont
  {Pawlowski}}]{braun_0908.0008}%
  \BibitemOpen
  \bibfield  {author} {\bibinfo {author} {\bibfnamefont {J.}~\bibnamefont
  {Braun}}, \bibinfo {author} {\bibfnamefont {L.~M.}\ \bibnamefont {Haas}},
  \bibinfo {author} {\bibfnamefont {F.}~\bibnamefont {Marhauser}}, \ and\
  \bibinfo {author} {\bibfnamefont {J.~M.}\ \bibnamefont {Pawlowski}},\
  }\href@noop {} {\bibfield  {journal} {\bibinfo  {journal} {Phys. Rev. Lett.}\
  }\textbf {\bibinfo {volume} {106}},\ \bibinfo {pages} {022002} (\bibinfo
  {year} {2011})},\ \Eprint {http://arxiv.org/abs/arXiv:0908.0008}
  {arXiv:0908.0008} \BibitemShut {NoStop}%
\bibitem [{\citenamefont {Bowman}\ and\ \citenamefont
  {Kapusta}(2009)}]{bowman09:_critical}%
  \BibitemOpen
  \bibfield  {author} {\bibinfo {author} {\bibfnamefont {E.~S.}\ \bibnamefont
  {Bowman}}\ and\ \bibinfo {author} {\bibfnamefont {J.~I.}\ \bibnamefont
  {Kapusta}},\ }\href@noop {} {\bibfield  {journal} {\bibinfo  {journal} {Phys.
  Rev. C}\ }\textbf {\bibinfo {volume} {79}},\ \bibinfo {pages} {015202}
  (\bibinfo {year} {2009})}\BibitemShut {NoStop}%
\bibitem [{\citenamefont {Skokov}\ \emph {et~al.}(2010)\citenamefont {Skokov},
  \citenamefont {Friman}, \citenamefont {Nakano}, \citenamefont {Redlich},\
  and\ \citenamefont
  {Schaefer}}]{skokov10:_vacuum_fluct_and_therm_of_chiral_model}%
  \BibitemOpen
  \bibfield  {author} {\bibinfo {author} {\bibfnamefont {V.}~\bibnamefont
  {Skokov}}, \bibinfo {author} {\bibfnamefont {B.}~\bibnamefont {Friman}},
  \bibinfo {author} {\bibfnamefont {E.}~\bibnamefont {Nakano}}, \bibinfo
  {author} {\bibfnamefont {K.}~\bibnamefont {Redlich}}, \ and\ \bibinfo
  {author} {\bibfnamefont {B.-J.}\ \bibnamefont {Schaefer}},\ }\href@noop {}
  {\bibfield  {journal} {\bibinfo  {journal} {Phys. Rev. D}\ }\textbf {\bibinfo
  {volume} {82}},\ \bibinfo {pages} {034029} (\bibinfo {year}
  {2010})}\BibitemShut {NoStop}%
\bibitem [{\citenamefont {McLerran}\ \emph {et~al.}(2009)\citenamefont
  {McLerran}, \citenamefont {Redlich},\ and\ \citenamefont
  {Sasaki}}]{mclerran09:_quark_matter_and_chiral_symmet_break}%
  \BibitemOpen
  \bibfield  {author} {\bibinfo {author} {\bibfnamefont {L.}~\bibnamefont
  {McLerran}}, \bibinfo {author} {\bibfnamefont {K.}~\bibnamefont {Redlich}}, \
  and\ \bibinfo {author} {\bibfnamefont {C.}~\bibnamefont {Sasaki}},\
  }\href@noop {} {\bibfield  {journal} {\bibinfo  {journal} {Nucl. Phys.}\
  }\textbf {\bibinfo {volume} {824}},\ \bibinfo {pages} {86} (\bibinfo {year}
  {2009})}\BibitemShut {NoStop}%
\bibitem [{\citenamefont {McLerran}\ and\ \citenamefont
  {Pisarski}(2007)}]{mclerran07:_phases_of_dense_quark_at}%
  \BibitemOpen
  \bibfield  {author} {\bibinfo {author} {\bibfnamefont {L.}~\bibnamefont
  {McLerran}}\ and\ \bibinfo {author} {\bibfnamefont {R.~D.}\ \bibnamefont
  {Pisarski}},\ }\href@noop {} {\bibfield  {journal} {\bibinfo  {journal}
  {Nucl. Phys.}\ }\textbf {\bibinfo {volume} {796}},\ \bibinfo {pages} {83}
  (\bibinfo {year} {2007})}\BibitemShut {NoStop}%
\bibitem [{\citenamefont {Fischer}\ and\ \citenamefont
  {Mueller}(2009)}]{fischer09:_chiral_dyson_schwin}%
  \BibitemOpen
  \bibfield  {author} {\bibinfo {author} {\bibfnamefont {C.~S.}\ \bibnamefont
  {Fischer}}\ and\ \bibinfo {author} {\bibfnamefont {J.~A.}\ \bibnamefont
  {Mueller}},\ }\href@noop {} {\bibfield  {journal} {\bibinfo  {journal} {Phys.
  Rev. D}\ }\textbf {\bibinfo {volume} {80}},\ \bibinfo {pages} {074029}
  (\bibinfo {year} {2009})}\BibitemShut {NoStop}%
\bibitem [{\citenamefont {Bilgici}\ \emph {et~al.}(2010)\citenamefont
  {Bilgici}, \citenamefont {F.~Bruckmann}, \citenamefont {Gattringer},
  \citenamefont {Hagen}, \citenamefont {Ilgenfritz},\ and\ \citenamefont
  {Maas}}]{bilgici10:_fermion_qcd}%
  \BibitemOpen
  \bibfield  {author} {\bibinfo {author} {\bibfnamefont {E.}~\bibnamefont
  {Bilgici}}, \bibinfo {author} {\bibfnamefont {J.~D.}\ \bibnamefont
  {F.~Bruckmann}}, \bibinfo {author} {\bibfnamefont {C.}~\bibnamefont
  {Gattringer}}, \bibinfo {author} {\bibfnamefont {C.}~\bibnamefont {Hagen}},
  \bibinfo {author} {\bibfnamefont {E.~M.}\ \bibnamefont {Ilgenfritz}}, \ and\
  \bibinfo {author} {\bibfnamefont {A.}~\bibnamefont {Maas}},\ }\href@noop {}
  {\bibfield  {journal} {\bibinfo  {journal} {Few Body Systems}\ }\textbf
  {\bibinfo {volume} {47}},\ \bibinfo {pages} {125} (\bibinfo {year} {2010})},\
  \Eprint {http://arxiv.org/abs/arXiv:0906.3957} {arXiv:0906.3957} \BibitemShut
  {NoStop}%
\bibitem [{\citenamefont {Kashiwa}\ \emph {et~al.}(2009)\citenamefont
  {Kashiwa}, \citenamefont {Kouno},\ and\ \citenamefont
  {Yahiro}}]{kashiwa09:_dual_polyak_nambu_jona_lasin}%
  \BibitemOpen
  \bibfield  {author} {\bibinfo {author} {\bibfnamefont {K.}~\bibnamefont
  {Kashiwa}}, \bibinfo {author} {\bibfnamefont {H.}~\bibnamefont {Kouno}}, \
  and\ \bibinfo {author} {\bibfnamefont {M.}~\bibnamefont {Yahiro}},\
  }\href@noop {} {\bibfield  {journal} {\bibinfo  {journal} {Phys. Rev. D}\
  }\textbf {\bibinfo {volume} {80}},\ \bibinfo {pages} {117901} (\bibinfo
  {year} {2009})}\BibitemShut {NoStop}%
\bibitem [{\citenamefont {Mukherjee}\ \emph {et~al.}(2010)\citenamefont
  {Mukherjee}, \citenamefont {Chen},\ and\ \citenamefont
  {Huang}}]{mukherjee:_chiral_polyak_nambu_jona_lasin}%
  \BibitemOpen
  \bibfield  {author} {\bibinfo {author} {\bibfnamefont {T.~K.}\ \bibnamefont
  {Mukherjee}}, \bibinfo {author} {\bibfnamefont {H.}~\bibnamefont {Chen}}, \
  and\ \bibinfo {author} {\bibfnamefont {M.}~\bibnamefont {Huang}},\
  }\href@noop {} {\bibfield  {journal} {\bibinfo  {journal} {Phys. Rev. D}\
  }\textbf {\bibinfo {volume} {82}},\ \bibinfo {pages} {034105} (\bibinfo
  {year} {2010})},\ \Eprint {http://arxiv.org/abs/1005.2482} {1005.2482}
  \BibitemShut {NoStop}%
\end{thebibliography}
\end{document}